\documentclass{article}
\usepackage{PRIMEarxiv}
\usepackage[colorlinks=true,linkcolor=red,urlcolor=blue,citecolor=blue,bookmarks=true]{hyperref}
\usepackage[utf8]{inputenc} 
\usepackage{lipsum}
\usepackage{amsfonts}
\usepackage{subfig}
\usepackage{graphicx}
\usepackage{epstopdf}
\usepackage{amsmath}
\usepackage{amssymb}
\usepackage{cleveref}
\usepackage{amsopn}
\usepackage{setspace}
\usepackage{color}
\usepackage{algorithmicx}
\usepackage[ruled]{algorithm}
\usepackage{algpseudocode}
\usepackage{algpascal}
\usepackage{threeparttable, tablefootnote}
\usepackage[table]{xcolor}
\usepackage{amsthm}
\usepackage{csquotes}
\usepackage{latexsym}
\usepackage[mathscr]{euscript}

\DeclareMathAlphabet\mathbfcal{OMS}{cmsy}{b}{n}

\usepackage[T1]{fontenc}    
\usepackage{url}            
\usepackage{booktabs}        
\usepackage{nicefrac}       
\usepackage{microtype}      
\usepackage{fancyhdr}       
\graphicspath{{media/}}     

\providecommand{\JEL}[2]
{
  \small	
  \textbf{\textit{JEL Codes:}} #2
}

{\title{Macroeconomic Predictions using Payments Data and Machine Learning\thanks{The opinions expressed herein are those of the authors and do not necessarily reflect those of the Bank of Canada. We thank Poclaire Kenmogne for his excellent assistance on this project. We also thank Alessandro Bitetto, Dalibor Stevanovic, Ingomar Krohn, Jonathan Chiu, Narayan Bulusu, Nicolas Woloszko, Pierre-Yves Yanni, Segun Bewaji, and Zhentong Lu for their comments. In addition, we thank discussants and participants of the following conferences for their comments and suggestions: IAAE Annual Conference (2021), 7th RCEA Time Series Workshop, Big Data and Machine Learning in Finance Conference (2021), 24th Central Bank Macroeconomic Modelling Workshop, Advanced Analytics: New Methods and Applications for Macroeconomic Policy Conference (2021), Time Series in Finance Workshop at ICAIF (2021), Conference on Non-traditional Data, Machine Learning and Natural Language Processing in Macroeconomics (2021), and the seminar at the Bank of Canada. This paper is accepted to present at the AEA 2023 Annual Meeting. $^\dagger$Corresponding author: ajit.ndesai@gmail.com}}}

\author{
  James~T.E. Chapman and Ajit Desai$^\dagger$  \\
  Bank of Canada \\
  Ottawa, ON, Canada \\
}

\begin{document}
\maketitle

\begin{abstract}
Predicting the economy's short-term dynamics---a vital input to economic agents' decision-making process---often uses lagged indicators in linear models. This is typically sufficient during normal times but could prove inadequate during crisis periods. 
This paper aims to demonstrate that non-traditional and timely data such as retail and wholesale payments, with the aid of nonlinear machine learning approaches, can provide policymakers with sophisticated models to accurately estimate key macroeconomic indicators in near real-time. Moreover, we provide a set of econometric tools to mitigate \emph{overfitting} and \emph{interpretability} challenges in machine learning models to improve their effectiveness for policy use. Our models with payments data, nonlinear methods, and tailored cross-validation approaches help improve macroeconomic nowcasting accuracy up to 40\%---with higher gains during the COVID-19 period. We observe that the contribution of payments data for economic predictions is small and linear during low and normal growth periods. However, the payments data contribution is large, asymmetrical, and nonlinear during strong negative or positive growth periods. 
\end{abstract}

\JEL J{C53, C55, E37, E42, E52}

\section{Introduction}

Consumers are increasingly adopting electronic payments; this has dramatically accelerated due to the COVID-19 pandemic~\cite{Paturi2020pc}. In the process, vast amounts of data have been generated. Much of this data is available in nearly real time.
Concurrently, recent advances in the field of machine learning (ML) provide a set of advanced econometric tools to analyze non-traditional data and nonlinear relationships, bringing new opportunities to efficiently process large-scale payments data. Thus, our objective in this paper is to demonstrate the usefulness of payments data and ML models to predict the economy’s short-term dynamics---known as nowcasting.

Predicting the economy’s short-term dynamics is a vital input into every economic agent's decision-making process. However, macroeconomic nowcasting is difficult for several reasons. For instance, many different data series are needed to describe the state of the economy adequately, but many of these data series, particularly official national account statistics, are released with significant lags~\cite{giannone2008nowcasting,angelini2011short}. 
This problem is especially difficult during times of crisis, such as the 2008 global financial crisis (GFC) and the COVID-19 pandemic, primarily because of the large and nonlinear economic impact of crises and the unconventional policy responses needed for their mitigation ~\cite{spange2010can,Hamilton20111006,greenwood2020predictable}. During such times, traditional models are inadequate because realizations of
target variables are far from their average values~\cite{vrontos2020modeling,coulombe2021can}. 

To address such challenges, econometricians have either used new data or developed new techniques~\cite{giannone2008nowcasting,choi2012predicting,buono2017big,bok2018macroeconomic,kapetanios2018big,koop2019macroeconomic,foroni2020forecasting}. We combine both new data and ML approaches to create a nowcast of the Canadian economy. First, we use comprehensive and timely settlement data from Canada's retail and large-value electronic payments systems. We then use the following five ML models: elastic net, support vector machines, random forest, gradient boosting, and artificial neural network~\cite{trevor2009elements}. We employ these parametric and non-parametric ML models because they are popular among time series forecasters and preferred in macroeconomic prediction problems~\cite{ahmed2010empirical,bok2018macroeconomic,athey2019ml}. 

ML models could prove useful in efficiently handling a wide variety of payments data and effectively managing collinearity in such data. This is beneficial because some of the payment streams used here are strongly correlated with each other~\cite{chapman2021covid}.
ML models can also help capture sudden, large, and possibly nonlinear effects of economic crises and the impact of unconventional policies designed to alleviate them~\cite{vrontos2020modeling,coulombe2021can}. 
This is important because different crises have reflected differently in payment streams, suggesting a tangled and possibly nonlinear relationship between some payments streams and macroeconomic targets.\footnote{In April 2020, the Canadian government began providing social benefits to citizens directly affected by COVID-19. This is reflected by a large increase in payment flows in the government direct deposit stream. Such a policy was not implemented during the GFC, yielding a drop in payment flows in this stream (see~\autoref{fig:macroVpayment}).} 
Moreover, ML models are beneficial when the emphasis is on improving prediction accuracy---a focus of the present paper~\cite{mullainathan2017machine,athey2017impact,yoon2021forecasting}.

The use of ML models, however, leads to many challenges that could reduce the effectiveness of these models for policy use. In particular, we mitigate the nowcasting ML model problem of overfitting, that is, due to the flexibility of these models, it is easy to overfit them on in-sample data, which could reduce their out-of-sample performance~\cite{bergmeir2012use,bergmeir2018note}. In addition, we address the difficulty in interpreting these models. Interpretability is important to understand their predictions---especially if they are used to support policy decisions~\cite{varian2014big,mullainathan2017machine,chakraborty2017machine,athey2019ml}.

To alleviate the classic ML issue of overfitting, we devise an improved cross-validation strategy tailored to macroeconomic prediction models. In cases where the out-of-sample test set has an economic crisis but the validation set, i.e., a part of the training set used for ML model tuning and cross-validation (see \autoref{fig:TimeSeriesSplit_validation}), does not, the traditional $k$-fold or leave-$p$-out cross-validation~\cite{trevor2009elements,bergmeir2012use} could be challenging because: (a) the standard $k$-fold splitting breaks the order (serial correlation) of the series,  
(b) the distribution of test and validation sets could differ, and (c) the model tuned predominantly on normal periods might not perform well on the out-of-sample crisis period. 
To overcome this, similar to~\cite{kuhn2013applied}, we use a randomized expanding window approach with $k$-fold cross-validation but without changing the order of the data. Since we have the COVID-19 crisis period in the test set, using random sampling helps to include a few samples from the GFC period in the validation set. Consequently, the distribution of validation and test sets are somewhat similar (see \autoref{fig:gdp_cv_sample_distributions} in \autoref{apn:cross_validation}), which could assist in selecting a model that performs well in both normal and crisis periods.  

Next, we address the interpretability issue by using the
SHapley Additive exPlanations (SHAP) methodology~\cite{lundbergNips2017unified,lundberg2020local2global}, based on Shapley values from the coalition game theory~\cite{shapley1953value, osborne1994gametheory}. 
To utilize this approach, we need to consider each nowcasting exercise as a \emph{game}. Shapley values can then be used to fairly distribute the \emph{payout} (i.e., the model prediction) among the \emph{players} (i.e., the predictors) of the game. 
SHAP provides a way to interpret ML model predictions at each nowcasting horizon in terms of the marginal contribution of each predictor toward the final prediction. Further, by averaging each prediction instance's contribution---in terms of Shapley values---we can compute the marginal contribution of each predictor for the entire sample. A similar approach is employed in the recent paper by~\cite{buckmann2021opening} for macroeconomic forecasting.

We observe that retail and large-value payments system data in the ML models---especially nonlinear gradient boosting regression (GBR)---can lower nowcast errors significantly. We obtain a 35--40$\%$ reduction in root-mean-square error (RMSE) in nowcasting GDP, retail trade sales (RTS), and wholesale trade sales (WTS)\footnote{We nowcast GDP because it is a crucial indicator for policymakers and commonly used to test nowcasting model performance. We nowcast RTS and WTS because we presume payments data have value in predicting them. Also, having multiple targets allows us to test the robustness of our models. Note: In Canada, all three target indicators are available at monthly frequencies and they are released with about a two months’ delay.} over a linear benchmark model.\footnote{As a benchmark model, we use the following series in our linear regression model: consumer price index (CPI), unemployment (UNE), Canadian financial stress indicator (CFSI), and the Conference Board's consumer confidence index (CBCI). Unemployment incorporates the effects of public sector hiring, and CPI is useful since we are using nominal predictors~\cite{galbraith2018nowcasting}. CFSI is a composite measure of systemic financial market stress for Canada~\cite{duprey2020canadian}. CBCC is based on a survey of Canadian households and has been shown to be useful in predicting household spending in Canada~\cite{kwan2006usefulness}.} 
Further, in the presence of payments data and ML models, compared to the dynamic factor models---commonly preferred for macroeconomic prediction---can reduce nowcasting RMSE by as much as 20--25\%. Out-of-sample performance gain using payments data and ML models is relatively greater (15 to 20\%) during the COVID-19 crisis period than the pre-COVID normal economic growth period, and the models using payments data perform better against the latest vintages compared to  the real-time vintages.

We also observe improved model performance when the proposed randomized expanding window approach with $k$-fold cross-validation is used for ML model tuning. The average RMSE across $k$-folds is 10--15\% smaller using the proposed  approach compared to the traditional expanding window $k$-fold cross-validation approach. 

Further, the Shapely value-based nowcasting model interpretations reveal that, in general, many payment streams are important along with other traditional predictors in nowcasting GDP, RTS, and WTS. Moreover, during the COVID-19 crisis period, the contribution of those payments streams is much higher than the benchmark predictors. 
Our analysis also suggests that the contribution of payments data in terms of Shapley values is small and linear during periods of low and normal growth. However, during periods of strong negative or positive growth, the payments data contribution is asymmetrical and nonlinear.\footnote{Contributions in terms of Shapley values from strong negative growth rates in payment streams are much stronger than similar values of positive growth rates (\autoref{fig:dependenceplots} and \ref{fig:dependenceplotsRTSWTS}).}

In summary, this paper demonstrates that combining timely data, nonlinear methods, tailored cross-validation approaches, and model agnostic interpretability tools can provide policymakers with sophisticated models to accurately estimate key macroeconomic indicators in near real-time, which is important to monitor the economy---especially during the crisis periods such as COVID-19.

In the past---driven by the need to overcome dependence on lagged variables---econometricians have used payments data for macroeconomic predictions~\cite{carlsen2010dankort,barnett2016nowcasting,duarte2017mixed,galbraith2018nowcasting,aprigliano2019using}. Canadian payments data are a particularly good candidate for nowcasting because they record transactions processed in various payment instruments. Thus, they capture a broad range of Canadian consumers, firms, and government economic activities. Also, these data are gathered electronically and hence are immediately available, and they are free of measurement or sampling errors~\cite{galbraith2007electronic}. 
Such datasets are shown to be useful during economic crisis periods such as the GFC and the COVID-19 shock~\cite{chetty2020did,bounie2020consumers,carvalho2020tracking,dahlhaus2021payment}. 

Traditionally, researchers have used data from a few selected payment instruments for nowcasting~\cite{galbraith2018nowcasting,aprigliano2019using}. One issue with this approach is that the use and importance of particular payment instruments may rise or fall for both economic and non-economic reasons.\footnote{In Canada, the proportion of electronic means of payment is increasing, and the use of cash is declining, primarily due to ease of accessibility driven by technological advancements. For instance, compared to 2018, the share of debit card payments processed through the Automated Clearing Settlement System (ACSS) increased by 21\%, and cash payments declined by 27\% in 2019~\cite{Paturi2020pc}.} Using data from one or two payment streams in isolation might not capture the full economic picture; therefore, in \cite{chapman2021covid}, the authors use most of the stream settled in Canada's retail payment system for macroeconomic nowcasting at the onset of COVID-19.
In this paper, however, we also include settlement data from Canada's high-value payments system and cover the wider COVID-19 period.\footnote{We use Large-Value Transfer System (LVTS) data, and it is among the top five contributors in nowcasting GDP~see \autoref{fig:summaryPlotAll}). Our out-of-sample testing period covers until December 2020.} 
Further, this paper addresses the ML models' interpretability issues and implements an improved cross-validation technique to overcome overfitting challenges. Additionally, this paper compares the performance of the nowcasting models in both normal and crisis periods for the target variables available at both real-time vintages and the latest vintages.

Recently---driven by the need to exploit non-traditional, complex, and large-scale datasets---econometricians have begun using ML models for macroeconomic nowcasting~\cite{chakraborty2017machine,richardson2020nowcasting,maehashi2020macroeconomic,chapman2021covid}. The cited articles suggest that ML models complement traditional econometric tools and are useful in extracting economic value from non-traditional data sources. Also, they show that in nowcasting, ML models often outperform traditional modeling approaches, such as ordinary least-squares and dynamic-factor models. 
However, in these papers, the interpretability and overfitting issues of ML nowcasting models are not adequately addressed, reducing the effectiveness of such models.

This paper proceeds as follows. Section \ref{sec:data} describes the payments systems data and discusses the adjustments performed on these data for macroeconomic predictions. Section \ref{sec:methods} provides a brief overview of various methods employed for nowcasting along with a discussion of challenges associated with using ML models for predictions. This is followed by a discussion of our results, in Section \ref{sec:results}. Finally, in Section \ref{sec:conclusion}, we set forth our conclusions. Several appendices provide further details on the payments data and the nowcasting methodology employed.



\section{Payments Systems Data}\label{sec:data}
The vast majority of non-cash transactions require settlement to extinguish the debt from the buyer to the seller. In modern economies, this is accomplished via centralized payments systems. The data coming from such systems are potentially useful because they are (a) timely, i.e., available immediately after the end of each period, (b) available at high-frequency, i.e., at the transaction or day levels, (c) precise, i.e., carry no sampling or measurement error, and (d) comprehensive, i.e., capture a broad range of financial activities across the country~\cite{galbraith2007electronic,chapman2021covid}. 

In Canada, the ACSS and LVTS are used to settle most transactions.\footnote{The ACSS supports $99\%$ of the daily transaction volume and $13\%$ of the daily value processed by Canadian payment systems. The LVTS settles $87\%$ of the total value moving through Canadian payment systems.} Our data consist of all settled transactions in both the ACSS and LVTS payments systems.
The ACSS settles the majority of retail and small-value payment items on a net basis. In 2019, the ACSS handled an average of 33 million transactions per business day, with an average daily total value of CA\$29 billion. 
The ACSS processes 22 payment streams. Broadly, these streams can be categorized into two groups: (1) electronic streams, which include, e.g., automated funds transfer (AFT), point-of-sale (POS) payments, and government direct deposit (GDD), and (2) paper streams, which incorporate encoded paper, paper remittances, and government paper items. 

In the ACSS, electronic means of payment have become more common than paper items due to their usability. This change is driven primarily by technological advancements leading to the inception and adoption of new payment instruments. However, economic crises such as the GFC and the COVID-19 shock also influence payment flows.
Historically, the encoded paper stream has the highest-value shares in the ACSS, followed by AFT credit. The POS payments stream has the largest volume of shares, followed by the encoded paper stream (see \cite{chapman2021covid} for the breakdown of shares of payment streams in the ACSS.) 

The LVTS facilitates the transfer of large-value payments between Canadian financial institutions on a gross basis. In 2019, the LVTS handled an average of approximately 40,000 transactions per business day, with an average daily value of CA\$189 billion. 
The LVTS provides each participant with two options called tranches, T1 and T2, to exchange payments. Each tranche (henceforth also referred to as a stream) differs based on how individual payments are collateralized.
Payments in the LVTS comprise foreign exchange payments, payments for settlement of Canadian-dollar-denominated securities, payments related to the final settlement of ACSS and Government of Canada transactions, as well as the Bank of Canada's own and its clients' payments. 
In the LVTS, payment value and volume are mostly processed through T2. Historically, T2 has processed roughly 75\% of the value and 98.7\% of the volume of payments, and T1 has processed roughly 25\% of the value and 1.3\% of the volume.

\subsection{Adjustments to Payments Data}
In the past, driven by technological advancements, some payment instruments from the ACSS were discontinued or merged into others, and several new payment instruments were created. 
For example, starting in 2012, a new stream was created to process the Government of Canada's direct deposit payments. This addition caused a sudden drop in the value and volume of payments in the AFT credit stream, where they were originally processed (see~\autoref{apn:payments_det} for specifics on changes in multiple ACSS streams over time).
To overcome the effects of such sudden changes and to get a better representation of payment flow, we merged several streams belonging to similar categories and settled related payments. 
Also, to overcome the effects of consumers' payment choices, i.e., when they switch payment method, because, for nowcasting, we are interested in capturing whether spending (or earning) has slowed (or stopped). We include the sum of all payment instruments in the ACSS ``Allstream" as a separate series. This should help develop an overall picture from the ACSS and mitigate the effects of unused streams (see \autoref{tab:payments_streams} footnotes for specifics on each adjustment performed).

After these adjustments, we are left with seven streams from the ACSS (which comprise transactions settled in all ACSS payment instruments) 
and two streams from the LVTS, which are listed in \autoref{tab:payments_streams} along with a short description.
For nowcasting, we use both the monthly gross dollar amount, i.e., {\it{value}}, and number of transactions, i.e., {\it{volume}}, settled in the payment instruments; this yields a total of 18 series. 

Like other macroeconomic time series, payments data have a strong seasonal component. We adjust all series (both value and volume) for seasonality using the X-13 ARIMA tool~\cite{x13seasonal}.
Note that recursive seasonal adjustments are performed in real time using the data available up to the nowcasting horizon at each time step.
Year-over-year (YOY) growth rates of the seasonality-adjusted payments series are used to predict the similarly adjusted YOY growth rates of macroeconomic indicators. Using growth rates (instead of levels) helps to induce (approximate) stationarity in both the target and predictors.

Our dataset does not include some payment instruments not settled through the ACSS or LVTS, such as credit card and e-transfer payments.\footnote{In 2019, credit card payments accounted for 6.2\% of the value and 31.1\% of the volume of total retail payments in Canada. Similarly, e-transfers accounted for 1.5\% of the value and 2.5\% of the volume~(\cite{Paturi2020pc}.} However, \cite{galbraith2018nowcasting} conclude that credit card payment data for Canada do not add significant value in nowcasting GDP and retail sales.\footnote{Note that in \cite{galbraith2018nowcasting}, the authors use a short sample size in their analysis of credit card data. The results could differ for a larger sample size.} Further, our dataset does not include {\it{on-us}} transactions where both sender and receiver have an account with the same financial institution; such transactions do not need to be settled in a payment system. However, their shares are small and may not materially influence our analysis.\footnote{On-us payments amount to roughly 20\% more than those settled in the ACSS. The value of on-us transactions differs by stream, for instance, in encoded paper, it is about 25\%, and in POS payments, it is 16\%~\cite{Paturi2020pc}.}

\rowcolors{2}{gray!10}{white}
\begin{table}[htbp]
  \begin{center}
  \begin{threeparttable}
    \caption{ACSS and LVTS payment streams used in this study\tnote{a}}
    \label{tab:payments_streams}
    \begin{tabular}{c l l}
      \toprule
      \textbf{ID} & \textbf{Stream} & \textbf{Short Description}\\
      \bottomrule
         C & AFT credit\tnote{b} & Government direct deposit (GDD): payrolls and account transfers\\
         D & AFT debit & Pre-authorized debit (PAD): automated bill and mortgage payments\\
         E & Encoded paper\tnote{c} & Paper bills of exchange: cheques, bank drafts, and paper PAD \\
         N & Shared ABM & Debit card payments to withdraw cash at shared ABM network \\
         P & POS payments\tnote{d} & Point-of-sale (POS) payments using debit card \\
         X & Corporate payments\tnote{e} & Exchange of corporate-to-corporate and bill payments \\
         All & Allstream\tnote{f} & The sum of all payment streams settled in the ACSS \\
         \hline
         T1 & LVTS-T1\tnote{g} & Time critical payments and payments to the Bank of Canada \\
         T2 & LVTS-T2\tnote{h} & Security settlement, foreign exchange, and other obligations \\
    \bottomrule
    \end{tabular}
    \begin{tablenotes}\footnotesize
        \item [a] The first six payment streams are representative of 20 payment instruments processed separately in the ACSS. There are a few additional payment instruments. However, they are not available for the entire period considered in this paper. Therefore, they are excluded from this study. The excluded streams are ICP regional image payments and ICP regional image payments return. Note: Excluded streams collectively account for only 0.001\% of the total value settled in the system. For further details on individual ACSS streams, see \autoref{apn:payments_det}.
        \item [b] Stream C is the sum of AFT credit and Government direct deposit streams. We combine them because, starting in April 2012, Government direct deposit was separated from the AFT credit stream and processed independently.
        \item [c] Stream E is the sum of multiple streams settled separately in the ACSS. It combines encoded paper (E), large-value encoded paper (L), image captured payments (O), Canada Savings Bonds (B), Receiver General warrants (G), and Treasury bills and bonds (H). It subtracts image-captured returns (S), unqualified (U), and computer rejects (Z) streams. We combine all of them because, over time, many of these streams were separated from the encoded paper stream and process similar types of payments. 
        \item [d] The value and volume of stream P are obtained by summing online payments (J) and POS payments (P) streams and subtracting online returns (K) and POS refunds (Q) streams. 
        \item [e] Stream X is the sum of paper remittances (F), EDI payments (X), and EDI remittances (Y). This stream is composed of all corporate-to-corporate payments and corporate bill payments and remittances.
        \item [f] Allstream is the sum of all payment streams processed in the ACSS.
        \item [g] We exclude payments from the Bank of Canada in stream T1.
        \item [h] The LVTS processes payment values equivalent to the annual GDP every five days, and the majority of the value and volume settled in the LVTS is processed in stream T2.
    \end{tablenotes}
    \end{threeparttable}
  \end{center}
\end{table}

\subsection{Payments Data for Macroeconomic Nowcasting}
The crux of the nowcasting problem is that most official estimates of macro indicators are released with a substantial delay. For instance, in Canada, GDP, RTS, and WTS are released with a delay of six to eight weeks.  
In addition, they undergo multiple revisions, sometimes years later, highlighting the uncertainty of their measurement.
Moreover, during a rapid crisis such as COVID-19, macroeconomic predictions are difficult because of the large and unprecedented economic impact.  
This can undermine the use of lagged data for nowcasting. 
Therefore, it is valuable to use more timely available information, in this case, payments systems data.

Payments data capture numerous types of transactions from both sides of macroeconomic accounts. For example, consumer income and expenditure, business-to-business payments, and Government of Canada spending. This variety, timeliness, and lack of sampling and measurement error in the payments dataset make it a rich economic information source.

For nowcasting exercises, we use Canada's monthly GDP, RTS, and WTS at the latest available vintages (i.e., after revisions) and real-time vintages (i.e., first release) as target variables.\footnote{Latest vintages of seasonally adjusted monthly GDP, RTS, and WTS are obtained from Statistics Canada Tables 36-10-0434-01, 20-10-0008-01, and 20-10-0074-01, respectively. Similarly, historical releases of GDP, RTS, and WTS are obtained from Tables 36-10-0491-01, 20-10-0054-01, and  20-10-0019-01, respectively.} We select these indicators because GDP is crucial for policymakers, and since we are using payments data, we think payments data have value in predicting RTS and WTS. All these indicators are released in Canada with a substantial lag and are available monthly for all historical releases. This variation allows us to test the robustness of our models. 

\begin{figure}[htbp]
 \centering
 \includegraphics[width=0.999\textwidth]{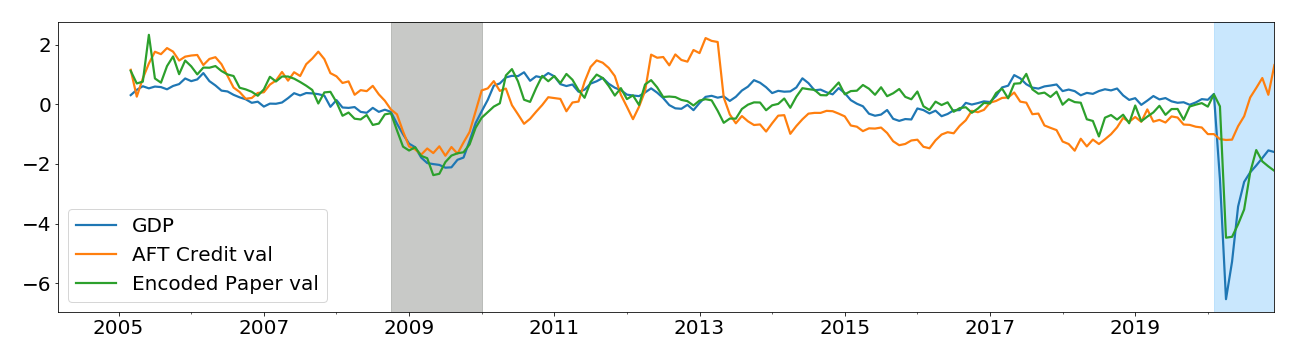}
 \includegraphics[width=0.999\textwidth]{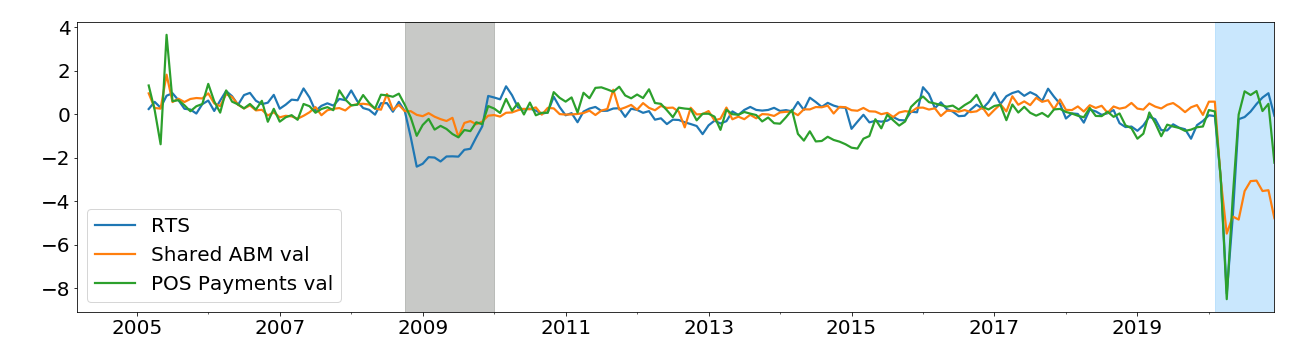} 
 \includegraphics[width=0.999\textwidth]{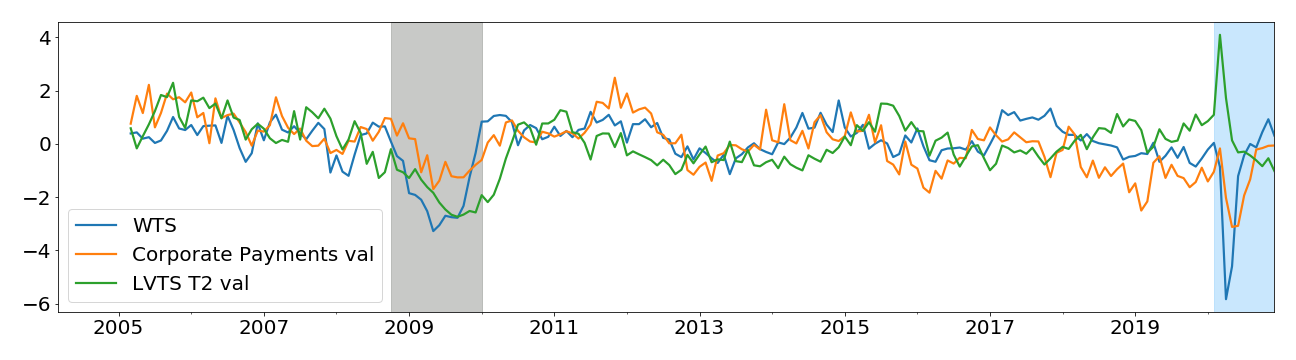}
 \caption{Standardized YOY growth rate comparisons of GDP, RTS, and WTS, with selected payment streams. Gray highlighting--GFC period; blue highlighting--COVID-19 period. Note: AFT credit includes Government direct deposit, encoded paper is the sum of multiple streams settled separately in the ACSS, POS payments include online payments, and corporate payments is the sum of paper remittances, EDI payments, and EDI remittances.} 
 \label{fig:macroVpayment}
\end{figure}

YOY growth rates of the latest monthly GDP are plotted with encoded paper and AFT credit values in \autoref{fig:macroVpayment} (top). Similarly, RTS's YOY growth rates are plotted with POS payments and shared ABM values in \autoref{fig:macroVpayment} (middle). The YOY growth rates of WTS are plotted with corporate payments and LVTS-T2 values in  \autoref{fig:macroVpayment} (bottom). To get a sense of the importance of payments data during a crisis, we highlight the growth rates of all variables during the 2008 GFC period (in gray) and the COVID-19 period (in blue).

During the GFC period, the decline and rebound in these payment streams' growth rates go hand-in-hand with macroeconomic indicators. Similarly, during the COVID-19 shock, we observe a sudden drop in most payment streams and in the macro variables. For instance, GDP and encoded paper, RTS and POS payments, along with WTS and corporate payments show similar movement during both crisis periods. This is a good indication of the economic value associated with these payment streams during such times.

During the COVID-19 period, however, we observe a complicated relationship between the macro indicators and some payment streams. This may be the result of the difference in the nature of the two crises. 
For instance, the value of payments through the AFT credit stream (which includes GDD payments) did not drop significantly at the onset of the COVID-19 shock. On the contrary, starting in April 2020, the value of payments processed through the AFT credit stream increased due to the flow of government social payments to those directly affected by the pandemic (\autoref{fig:macroVpayment}, top). Similarly, we note that the value of payments through the LVTS-T2 stream surged significantly at the onset of COVID-19, showing an opposite behavior to that of macro indicators during the same period. Such behavior is not seen during the GFC period, where both WTS and LVTS-T2 growth rates drop (\autoref{fig:macroVpayment}, bottom).\footnote{Similar behavior is observed in LVTS-T1, where payment value rose dramatically at the onset of COVID-19 due to extraordinary measures taken by the Bank of Canada under its quantitative easing policy~\cite{MPRapr2020}.} Such complex behavior could be challenging to capture using traditional linear models. 




\section{Methodology}\label{sec:methods}
This section briefly describes the nowcasting models employed. First, we discuss ordinary least squares (OLS) and the dynamic factor model (DFM). This is followed by a brief discussion of the ML models.  

Consider a set $X = \{{\bf{x}}^1, {\bf{x}}^2, \dots, {\bf{x}}^M\}$ of $M$ predictors (also called features or independent variables) and a target $\bf{y}$ (dependent variable), each with $N$ data (sample) points. For example, predictors could be monthly aggregated values settled in each payments stream, and the target could be monthly GDP (both recorded at the end of each month). 
This can be represented as a dataset $(X, \bf{y})$ where $X$ is of size $N \times M$ and ${\bf{y}}$ is a vector of size $N \times 1$. Let us denote $\hat{\bf{y}}$ as the predicted target, which can be estimated, for example, using payments data and OLS model as
\begin{eqnarray}\label{eqn:model}
  \hat{\bf{y}}(X, \bf{w}) =& X {\bf{w}},
\end{eqnarray}
where ${\bf{w}}$ is a vector of unknown coefficients (betas or weights) of size $M \times 1$. 
In OLS, the objective is to minimize the residual sum of squares between the observed ${\bf{y}}$ and the predicted $\hat{{\bf{y}}}$ target variables. 
Such linear models have proven to be valuable and straightforward for prediction, and they are commonly used due to their simplicity and interpretability. However OLS can model only linear relationships in the parameters $\bf{w}$. Although the linearity assumption makes them easy to interpret on a modular level, it generally does not perform well on wide, large, and complex datasets~\cite{trevor2009elements}. Moreover, multicollinearity in the predictors, although does not reduce the predictive power, could lead to reduced precision of the estimated coefficients. 

The DFMs are a powerful approach to capturing the common dynamics of a set of predictors in a relatively small number of latent factors. This can act as a dimension-reduction technique by estimating a small set of dynamic factors from a large set of observed variables.
DFM is a frequently selected model for macroeconomic nowcasting and forecasting~\cite{giannone2008nowcasting, stockDFM16}. 
Similar to~\cite{chernis2017dynamic}, we estimate the factors using the model of~\cite{banbura2014maximum}, which effectively handles a large number of predictors. The basic representation of the model is 
\begin{align}\label{eqn:dfm}
X_t &= \Lambda f_t + \epsilon_t \\
f_t &= A_1 f_{t-1} + \dots + A_p f_{t-p} + u_t,
\end{align}
where $X_t$ is a set of predictors at time $t$, $f_t$ is the unobserved factor at $t$, $\Lambda$ is the vector of factor loadings, $\epsilon_t$ is the idiosyncratic disturbance at $t$, $A_i$ are matrices of autoregression coefficients, and $u_t$ is the factor disturbance at $t$. 
DFMs are successfully applied for economic monitoring and predictions around the world~\cite{banbura2010nowcasting,stockDFM16,hindrayanto2016forecasting, bragoli2017now} including nowcasting Canada's GDP~\cite{chernis2017dynamic,chernis2020three}. Therefore, we employ this model to nowcast various Canadian macro indicators using payments data. 

\subsection{Machine Learning Models for Nowcasting}
To exploit non-traditional and large-scale data sources, researchers have recently begun utilizing ML models for economic nowcasting~\cite{richardson2020nowcasting,maehashi2020macroeconomic,chapman2021covid}. ML models have been shown to handle wide- and large-scale data efficiently and can manage collinearity. Further, they have been demonstrated to capture nonlinear interactions between the predictors and the target~\cite{chakraborty2017machine,yoon2021forecasting,coulombe2020machine,coulombe2021can,buckmann2021opening}.

We use some of the recently popularized parametric and non-parametric ML approaches, such as elastic net~\cite{zou2005regularization}, support vector machines~\cite{smola2004tutorial}, random forest~\cite{breiman2001random}, gradient boosting~\cite{friedman2001greedy}, and  feedforward  artificial  neural networks~\cite{bengio2009learning}. 
For each model, there are many variations proposed in the literature. However, we focus on the simpler version of each model to test their feasibility for macroeconomic nowcasting in Canada. In the remainder of this section, we provide a high-level description of these models. For further details on them, refer to \autoref{apn:model_details} and various textbooks~\cite{friedman2001elements,buckmann2021opening,trevor2009elements,bengio2009learning}.  

The elastic net (ENT) is a regularized linear regression model. Here, the objective is similar to that of the OLS with the addition of $L_1$ and $L_2$ penalties depending on how large the sum of the parameters $\bf{w}$ can become.\footnote{A regression model that uses only the $L_1$ penalty is a Lasso regression, and a model that uses only the $L_2$ penalty is a Ridge regression~\cite{trevor2009elements,zou2005regularization}.} In an ENT regression, the combination of $L_1$ and $L_2$ penalties allows for learning a sparse model while encouraging grouping effects, stabilizing regularization paths, and removing limitations on the number of selected variables~\cite{zou2005regularization}.

Support vector regression (SVR) is based on support vector machines, where the task is to find a hyperplane that separates the entire training dataset into, for example, two groups, by using a small subset of training points called support vectors. 
In SVR, the main objective is to determine a decision boundary at a distance from the support vectors such that the data points closest to the hyperplane are within that boundary line. This gives us the flexibility to define how much error is acceptable in our model~\cite{burges1998tutorial,smola2004tutorial}.

Another popular approach is random forest regression (RFR), a decision tree-based ensemble learning method built using a forest of many regression trees. This is a non-parametric approach that addresses the multicollinearity problem slightly differently from parametric approaches such as ENT.
Random forest is a bagging (bootstrap aggregation) approach, that is, each tree is independently built from a subset of the training dataset. Each sample randomly selects a subset of features from the available set of features, helping in decorrelation. The final prediction is performed by averaging the predictions of all regression trees~\cite{breiman2001random,liaw2002classification}.

Similar to RFR, gradient boosting regression (GBR) is a tree-based, non-parametric ensemble learning approach. However, unlike RFR, GBR is based on a boosting in which a sequence of weak learners (decision trees) are built upon a repeatedly modified version of the training dataset. In this approach, the base learners are sequentially improved by repeatedly applying the same base learner with the target's residuals as the outcome of interest~\cite{friedman2001greedy}.

The feedforward artificial neural network (ANN) with hidden layers consists of multiple layers of artificial neurons sandwiched between input and output layers.
In this approach, the data always move forward through the network layers. The weighted sum of the first layers is typically passed through a nonlinear activation function resulting in a nonlinear function of the inputs. Then the outputs are sent to the next layer, and the process continues until the last layer. Once we obtain the final output from the network, we measure how good that output is compared to the target's actual value using an objective function, for example, mean squared error. Given these results, we go back and adjust the weights and biases of the network. Typically we need a large training dataset to achieve good performance using ANN~\cite{bengio2009learning,goodfellow2016deep}.

Note that there are many advanced versions of both tree-based methods, such as LightGBM~\cite{ke2017lightgbm}, and deep ANN-based methods, such as long short-term memory (LSTM)~\cite{hochreiter1997long} proposed in the literature. However, to efficiently utilize them for prediction often requires a large training sample. Since our dataset is small (about 200 sample points), these models do not perform better than the models used in this paper.

\subsection{Machine Learning Model Cross-Validation}\label{sec:cv}
Overfitting is commonly attributed to the use of ML models---especially nonlinear models. ML models have many parameters that can be optimized to improve prediction accuracy (commonly known as hyperparameter tuning). Therefore, it is straightforward to tune the model to perform well on a specific dataset, for example, an in-sample training set. However, such models generally fail to perform well when applied to unseen data~\cite{trevor2009elements}.

This problem can be alleviated using $k$-fold cross-validation techniques. In the standard approach, the training sample is randomly split into $k$-folds, then for each iteration, the $k-1$ folds are used for in-sample training, and the $k^{th}$ fold is used for out-of-sample testing~\cite{trevor2009elements}. Such procedure effectively avoids overfitting. However, random splitting of the training sample breaks the order of series (autocorrelation) and could lead to the use of future data points for past predictions, giving an unfair advantage to the model. For these reasons, it may not practical to use it in the \emph{same way} in nowcasting models~\cite{bergmeir2012use}. 

\begin{figure}[htbp]
 \centering
 \includegraphics[width=0.99\textwidth]{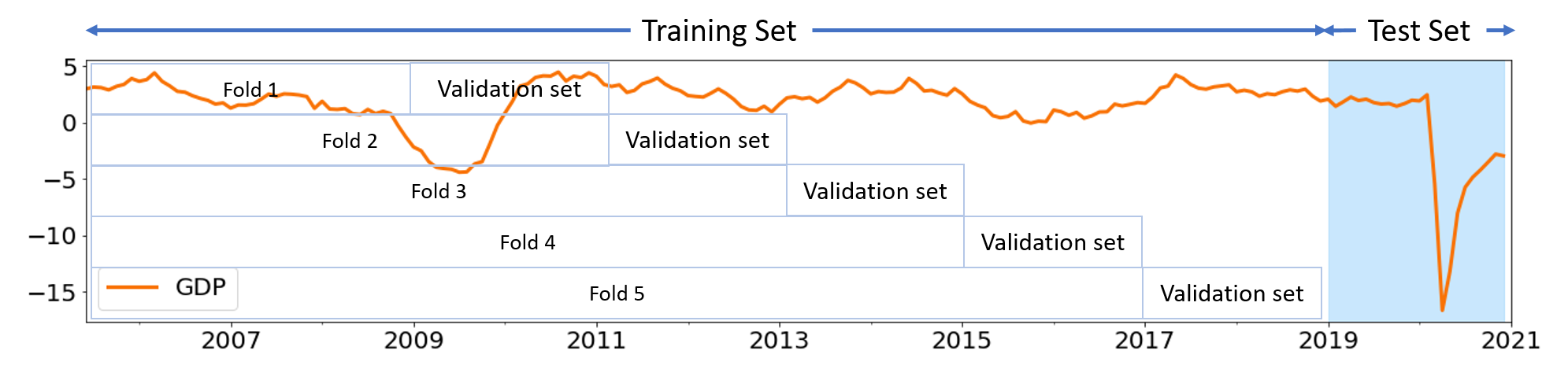}
 \includegraphics[width=0.99\textwidth]{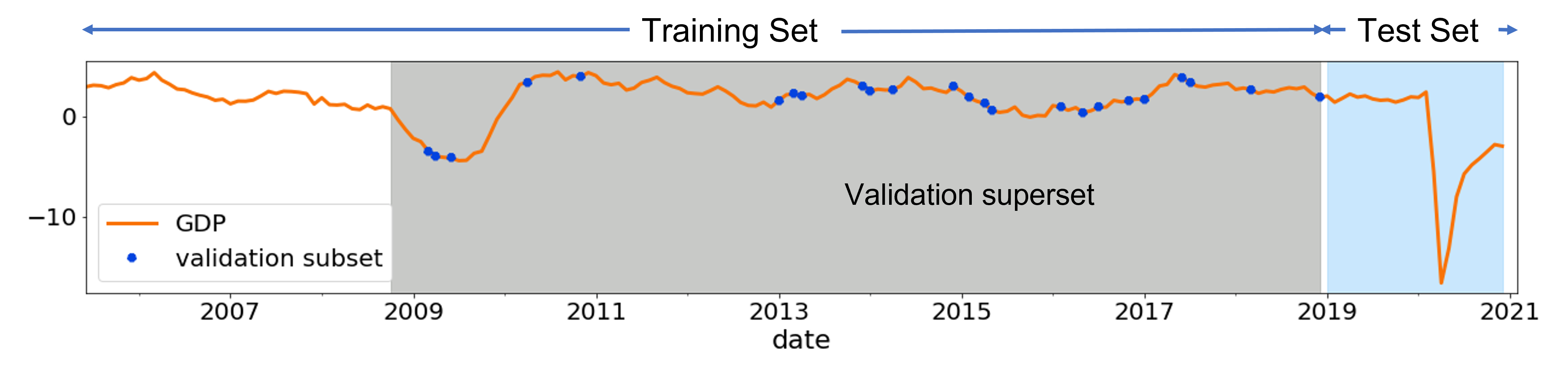}
  \caption{(Top) Schematic of standard expanding window approach for cross-validation in time series. The dataset is divided into a training set with validation subsets and a test set (highlighted in blue). (Bottom) Schematic of the proposed randomized expanding window approach showing a typical validation subsets (represented by {\color{blue} ${\bullet}$}) randomly sampled from the validation superset (highlighted in gray). In both plots, the orange line shows the GDP growth rate.}
 \label{fig:cv_expanding_window}
\end{figure}

\begin{figure}[htbp]
 \centering
 \includegraphics[width=0.999\textwidth]{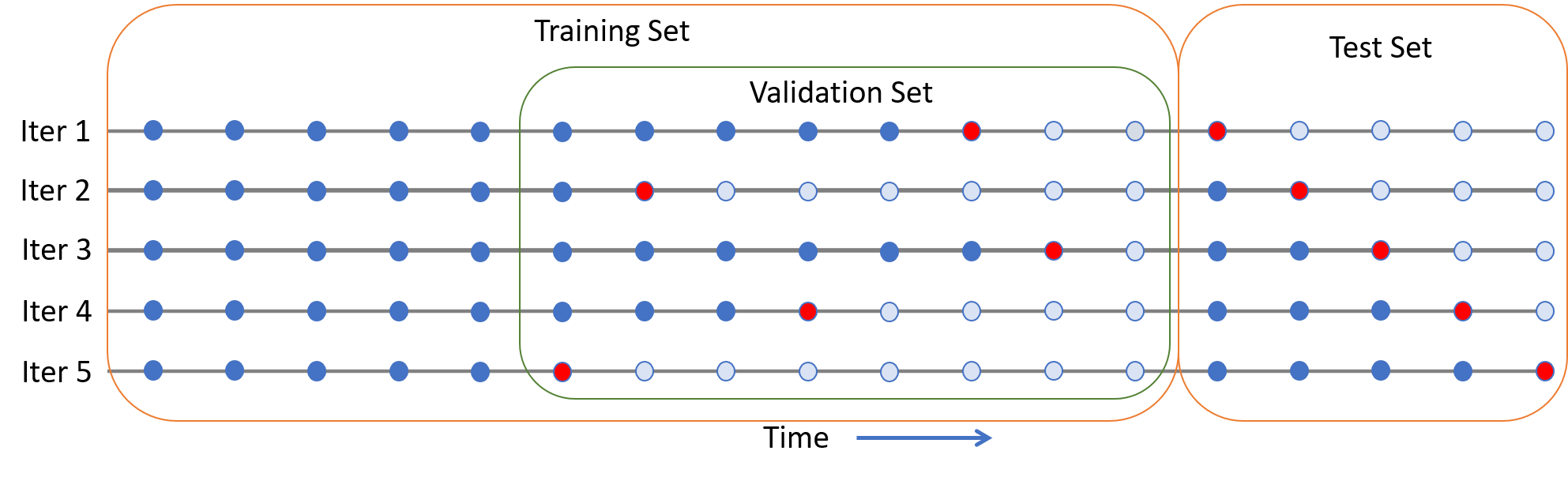}
  \caption{Schematic of expanding window approach for a typical fold in $k$-folds cross-validation and out-of-sample prediction. The available data are divided into training, validation, and test sets. For the given iterations of the expanding window (Iter), {\color{blue} ${\bullet}$} represents in-sample training points and {\color{red} ${\bullet}$} represents out-of-sample test points (for the fold). For each iteration in this fold of cross-validation, we use randomly sampled {\color{red} ${\bullet}$} points from the validation superset as the validation subset. Note: the out-of-sample size (the number of {\color{red} ${\bullet}$} points) in each validation subset is kept similar to the test set. For instance, both the validation subset and test set have five out-of-sample points each in this schematic.} 
 \label{fig:TimeSeriesSplit_validation}
\end{figure}

This challenge can be mitigated using an expanding window approach for cross-validation~(\autoref{fig:cv_expanding_window}, top). Here, the end portion of the training set, often called a validation set, is kept aside for model tuning and cross-validation. For each iteration of the expanding window, the training sample is increased by one period and the model prediction is performed on the next period from the validation sets~\cite{bergmeir2012use}.
Consequently, the model parameters can be chosen based on model performance on the validation sets. See~\autoref{apn:cross_validation} for additional details. 

Such approach is useful for nowcasting during normal economic periods. However, in cases where the test sample includes an economic crisis but the validation sample does not, traditional expanding window cross-validation could be challenging because (a) the distribution of test and validation samples are quite different and (b) the model is predominantly tuned for normal periods and therefore may not perform well for out-of-sample crisis periods. 

We implement a slightly altered version of the expanding window approach tailored to macroeconomic nowcasting models~ (\autoref{fig:cv_expanding_window}, bottom). 
We randomly sample $n$-points (one for each iteration) from the validation superset (highlighted in gray), beginning just before the GFC and continuing to the end of the training set, and use them as a validation sample. For each iteration of expanding window validation, only data points that come before the chosen point are used for training---preserving the order of data and temporal dependency between observations (see \autoref{fig:TimeSeriesSplit_validation}). 

In the current exercises, since we have the COVID-19 period in the test set, using a random sampling strategy leads us to include a few sample points from the GFC in each validation subset. Therefore, the proposed approach helps make the distribution of validation and test sets somewhat similar (see \autoref{fig:gdp_cv_sample_distributions} in \autoref{apn:cross_validation}) and assists in selecting a model that can perform well for both normal and crisis periods. Also, our approach removes the restriction on the number of validation sets we can sample that could be binding in traditional cross-validation approaches. However, in our approach, due to random sampling, some observations may be selected more than once in the validation subsets and some may never be selected. This could lead to overfitting if too many validation sets are sampled.

Further, instead of using all payment streams in each model or manually selecting a few payment streams for the given macro indicator, we use a \emph{data-driven} approach for feature selection. We treat the number of payment steams $p$ as similar to a model parameter and use the expanding window cross-validation approach to optimally select the best $p$ streams for each target variable based on their performance on the training and validation sets. Further details on cross-validation and model selection are discussed in~\autoref{apn:selectBestFeatures}.

\subsection{Machine Learning Model Interpretability}\label{sec:mlinterpret}

Another problem commonly attributed to the use of ML model is the loss of interpretability due to their complex nature. However, interpretability is essential for many applications---including macroeconomic prediction~\cite{mullainathan2017machine,athey2019ml,buckmann2021opening}.

Some ML models employed here, such as ENT, SVR, and the simple tree-based learning models, are inherently interpretable, but only to a certain extent~\cite{zou2005regularization,breiman2001random}. Tree-based ensemble approaches such as GBR and RFR can also be interpreted---to some extent---using impurity or permutation-based feature importance methods~\cite{breiman2001random, molnar2020interpretable}. However, they are mostly used for global (entire sample) interpretations. Also, each of those methods has different interpretability approaches, making them difficult to compare. To address these challenges, we use the Shapley value-based model agnostic approach SHAP developed in~\cite{lundbergNips2017unified}. The SHAP can be used for both the local (each prediction instance) and global interpretations. Moreover, the SHAP can be used for dependence plots which could be valuable to get further insights.

In SHAP, the Shapley value method from coalitional game theory~\cite{shapley1953value,osborne1994gametheory} 
is used to fairly distribute the \enquote{payout} (= the prediction) among the \enquote{players} (= the predictors)~\cite{lundberg2020local2global}.
In nowcasting, SHAP can be used to fairly distribute the ML model prediction among the set of predictors $X_t$ at each time horizon $t$ for local model interpretation. 
Further, using Shapley values for each instance $t$, we can compute the global interpretation of the ML models in the form of feature importance for the entire training or testing datasets.

\cite{lundbergNips2017unified} propose two approaches based on the type of underlying process used to compute the Shapley values: (1) KernelSHAP, a kernel-based estimation approach, which can be used for many ML models, such as ENT, ANN, and tree-based models, and (2) TreeSHAP, a computationally efficient approach for Shapley value estimation used only for tree-based ML models, such as decision trees, random forests, and gradient boosting models.

SHAP approaches are reliable because they are developed based on the Shapley value, which has game-theoretical foundations. However, the time required to estimate Shapley values could increase exponentially with the number of predictors. This is not a significant concern for our application because we have comparatively fewer predictors and smaller sample sizes. The KernelSHAP method also suffers from collinearity in the predictors, which could represent a concern for our work, given that several predictors are correlated. These problems can be mitigated to an extent using TreeSHAP, but only for tree-based models. 
Another challenge with SHAP approaches is that it is possible to create intentionally misleading interpretations to hide the bias. 
Also, in some cases, the outcomes can easily misinterpret and lead to ambiguous conclusions. Therefore, SHAP should be used with caution~\cite{alvarez2018robustness,slack2020fooling,molnar2020interpretable}.\footnote{Note that the Shapley value-based interpretation approach does not provide causal inference or any optimal statistical criterion. They only explain the marginal contribution of each feature to the difference between the actual prediction and the mean prediction given the set of predictors. The Shapley value-based approaches could also be used to select the best subset of predictors. However, in the context of model interpretation, the focus is only on computing the marginal contribution of \emph{all predictors} involved in the prediction exercise~\cite{lundbergNips2017unified,molnar2020interpretable}.} Further details on Shapley values and SHAP with a test example are given in~\autoref{apn:shapvalue}.

\subsection{Case Specifications and Model Training}\label{subsec:training}
As a benchmark (or base case), we employ a linear regression model using OLS. 
Here, we use the first available lagged target variable along with the latest available CPI, unemployment, CFSI,\footnote{CFSI is computed using data from the following seven market segments: equity market, Government of Canada bond market, foreign exchange market, money market, bank loans market, corporate bonds market, and housing market.} and CBCI.\footnote{CBCI is based on the Conference Board's survey of Canadian households, which provides a measure of consumer optimism on current economic conditions.} 
CFSI and CBCI are available immediately after period-end and carry comprehensive and useful information about the macro indicators. Along with CPI and UNE (available with a one-to-two week delay), these predictors form a strong benchmark to assess information gain using payments systems data.

In the main case of interest, along with the predictors specified in the base case, we use the payments data listed in \autoref{tab:payments_streams}. Here, we first use 
DFM to assess the marginal contribution of payments data when used in a sophisticated econometric model. 
Next, we test the usefulness of the various ML models discussed above in Section \ref{sec:methods}. Finally, we compare the performance of the ML models with the benchmark case and DFM. We use RMSE as the key performance indicator for out-of-sample model evaluation. 

For all cases, using a procedure similar to ~\cite{giannone2008nowcasting,galbraith2018nowcasting}, we perform nowcasting at three monthly time horizons, extending from the start of the month of interest ($t$) until the month before the official release ($t+2$). As we advance in time, we include new predictors when they become available.
For example, GDP nowcasting at time horizon $t$, that is, on the first day of the month of interest, we use the latest available benchmark variables and the monthly aggregated payments data available at $t-1$. Model $\mathcal{F}$ can be specified as\footnote{GDP is released with a two-month lag, CPI and UNE are released with a one-to-two week lag, and CFSI, CBCC, and payments data are available the day after the end of the period.} 
\begin{eqnarray}
    \widehat{GDP}_{t} =& \mathcal{F}({GDP}_{t-3}, \ {CPI}_{t-2}, \ {UNE}_{t-2}, \ {CFSI}_{t-1}, \ {CBCC}_{t-1}, \ {Payments}_{t-1}).
\end{eqnarray}
Similarly, at the next nowcasting horizons $t+1$ and $t+2$, using the latest available predictors, the models can be specified as\footnote{At the $t+2$ nowcasting horizon (on the first day of the month in which the target month's macro indicators are released), we have $t+1$ months of payments data. However, we do not include this because we are interested mainly in assessing the usefulness of $t$ month's payment data to predict $t$ month's macro variables. Also, note that the left-hand sides in Equations 4, 5, and 6 represent the same month's target but they are estimated at different time horizons.} 
\begin{eqnarray}
    \widehat{GDP}_{t+1} =& \mathcal{F}({GDP}_{t-2}, \ {CPI}_{t-1}, \ {UNE}_{t-1}, \ {CFSI}_{t}, \ {CBCC}_{t}, \ {Payments}_t), \\
     \widehat{GDP}_{t+2} =& \mathcal{F}({GDP}_{t-1}, \ {CPI}_{t}, \ {UNE}_{t}, \ {CFSI}_{t}, \ {CBCC}_{t}, \ {Payments}_t).
\end{eqnarray}

We train the nowcasting models using the expanding window approach as schematically outlined in~\autoref{fig:TimeSeriesSplit_validation}. First, we divide the dataset into two subsets: a training set (in-sample) for model training and a testing set (out-of-sample) for predictions. The OLS and DFM models are directly trained on the training set and used for predictions on the test set. The ML models, which require extensive hyperparameter tuning and cross-validation, are trained using the following procedure: 
\begin{enumerate}
\item From the training sample, we select two dates covering the wider range of training data as a validation superset (\autoref{fig:cv_expanding_window}) and randomly choose a set of $n$ sample points as a validation set (where $n=24$ points, it is the same size as the test sample).\footnote{We choose a start date just before the GFC period and an end date just before the test set, then select $n$ random data points between these two dates as a validation set. This helps to include a few data points from the crisis period in each fold of the validation subset, at the same time avoiding use of a large cross-validation sample.}
\item Thereafter, for each sample date in the validation subset, we select all the sample points before that date for training and use the sample date for prediction (\autoref{fig:TimeSeriesSplit_validation}). In this way, we maintain temporal dependency and avoid using future data for predictions in the past.
\item Next, for each model, we specify the grid for selected hyperparameters. Then, 
for each value of a specified parameter, we iterate over the validation subset and compute RMSE.
\item Steps 2 and 3 are repeated $k$ times for the same set of hyperparameters but with a different validation subset randomly sampled from the validation superset ($k=5$ fold cross-validation).  
\item Next, we select the best set of model parameters, i.e., the parameters with the lowest average validation RMSE (averaged over $k$ folds) as the final model. 
\item Finally, the chosen model is used for predictions on the test set by utilizing the standard expanding window approach over the training and test set (\autoref{fig:TimeSeriesSplit_validation}). 
\end{enumerate}

\section{Results and Discussion}\label{sec:results}
The payments data used for nowcasting exercises range from Mar 2004 to Dec 2020. The in-sample training period is Mar 2005 to Dec 2018 ($N=166$ sample points),\footnote{We lose the first full year of data after computing YOY growth rates.} and the out-of-sample testing period is Jan 2019 to Dec 2020 ($N=24$). Our training set includes the 2008 GFC period, and the test set combines a normal economic growth period (Jan 2019 to Feb 2020) and part of the ongoing COVID-19 crisis period (Mar 2020 to Dec 2020). This allows us to examine model performance during both normal and crisis periods.

YOY GDP, RTS, and WTS growth rate nowcasting performance for the various cases outlined in Section \ref{subsec:training} are discussed below. 
\autoref{tab:olsdfm} compares nowcasting performance---in terms of out-of-sample RMSE---of the DFM and ML model specifically, gradient boosting regression because it consistently performed better across different target variables and time-horizons compared to the other models (see \autoref{tab:dmfml} in \autoref{apn:model_details}). on the main case against the benchmark models at the time horizons $t$, $t+1$, and $t+2$.

\rowcolors{2}{gray!10}{white}
\begin{table}[htbp]
  \begin{center}
  \begin{threeparttable}
    \caption{Out-of-sample RMSE comparisons for seasonally adjusted YOY growth rate of macro variables at time horizon $t$ on the first day of the month of interest (top panel), $t+1$ on the first day after the month of interest (middle panel), and $t+2$ on the first day, two months after the month of interest (bottom panel)\tnote{a}}
    \label{tab:olsdfm}
    \begin{tabular}{c c c c c c}
      \toprule
      \addlinespace
       \textbf{Target}\tnote{b} \  & \ \textbf{Benchmark}\tnote{c} \ & \  \textbf{Main DFM}\tnote{d} \ & \ \textbf{Main ML}\tnote{e} \ & \ \textbf{RMSE Reduction ($\%$)}\tnote{f} \\
      \addlinespace
      \bottomrule
         GDP  & 4.58  & 3.95\tnote{} & 3.70 & 19 \\
         RTS  & 7.88  & 7.40\tnote{} & 7.38 & 7 \\
         WTS  & 6.34  & 5.81\tnote{} & 5.74 & 10 \\
      \bottomrule
      \end{tabular}
      \begin{tabular}{c c c c c c}
      \toprule
      \addlinespace
       \textbf{Target} \ & \ \textbf{Benchmark} \ & \  \textbf{Main DFM} \ & \ \textbf{Main ML\tnote{}} \ & \ \textbf{RMSE Reduction ($\%$)\tnote{}} \\
      \addlinespace
      \bottomrule
         GDP  & 3.97  & 2.98 & 2.43\tnote{*} & 39 \\
         RTS  & 8.47  & 6.36 & 5.44\tnote{*} & 36 \\
         WTS  & 7.17  & 6.18 & 4.28\tnote{*} & 41 \\
      \bottomrule
      \end{tabular}
      \begin{tabular}{c c c c c c}
      \toprule
      \addlinespace
       \textbf{Target} \ & \ \textbf{Benchmark} \ & \ \textbf{Main DFM} \ & \ \textbf{Main ML\tnote{}} \ & \ \textbf{RMSE Reduction ($\%$)\tnote{}} \\
      \addlinespace
      \bottomrule
         GDP  & 2.84 & 2.63\tnote{} & 2.18 & 23 \\
         RTS  & 7.60 & 6.15\tnote{} & 5.55 & 25 \\
         WTS  & 6.24 & 5.76\tnote{} & 4.72 & 24 \\
      \bottomrule
      \end{tabular}
    \begin{tablenotes}\footnotesize
    \item [a] In-sample training period, Mar 2005 to Dec 2018 ($p=166$), and out-of-sample testing period, Jan 2019 to Dec 2020 ($p=24$). 
    \item [b] GDP-gross domestic product, RTS-retail trade sales, WTS-wholesale trade sales. Note: we use the latest available values of these targets. We also perform similar exercises by using target variables at first release (real-time vintages). These results are presented in \autoref{apn:realtimevslatest}.
    \item [c] As a benchmark, we use OLS with CPI, UNE, CFSI, CBCC, and the first available lagged target variable (i.e., the second lag at nowcasting horizon $t$).
    \item [d] For the main DFM case, we use payments data along with the predictors in the benchmark case. Similar to the model employed in~\cite{chernis2017dynamic}, we use the DFM model with two factors and one lag in the VAR driving the dynamics of those factors. Idiosyncratic components are assumed to follow an AR(1) process. Note: including additional factors does not improve model performance. 
    \item [e] We use GBR because it consistently performs better than other ML models (see \autoref{tab:dmfml} in \autoref{apn:model_details}). We select model parameters using the cross-validation procedure outlined in~\autoref{apn:cross_validation} and~\ref{apn:selectBestFeatures}. For example, the selected model for GDP nowcasting at $t+1$: $learning\_rate$ is 0.1, $max\_depth$ is 1, and $n\_estimators$ is 1000 (see~\autoref{apn:model_details} for further details). 
    \item [f] Percentage reduction in RMSE over the benchmark model using ML on the main case.
    \item [*, **, ***] denote statistical significance at the 10, 5, and 1\% levels, respectively, for the Diebold-Mariano test using the benchmark.
    \end{tablenotes}
    \end{threeparttable}
  \end{center}
\end{table}

Our results suggest that the payments systems data in conjunction with ML models can provide notable reductions in nowcasting RMSEs for all three macro variables considered.  
Specifically, we observe a 35--$40\%$ reduction in RMSE over the benchmark case in nowcasting GDP, RTS, and WTS at time horizon $t+1$, i.e., when we use same month's payments data as the target variable. The main case predictions at this time horizon are statistically significant for the Diebold-Mariano test using the benchmark.\footnote{We recognize that the Diebold-Mariano test has poor finite-sample properties. However, we use it to be comparable with similar papers where it is used, e.g., ~\cite{chernis2017dynamic} and \cite{aprigliano2019using}.}

Comparatively, the information gain using payments data is smaller at nowcasting horizon $t$, that is, when we use the first lag of payments data, and $t+2$, that is, when the values of 
all benchmark indicators are available at $t$ along with the first lag of the target variables. 
In these cases, we obtain a 7--$25\%$ reduction in RMSE over the benchmark in nowcasting GDP, RTS, and WTS. These results suggest that payments data provide the greatest nowcasting value when the given month's payments data are used immediately to predict the same month's macro variables (at $t+1$ time horizon, i.e., on the first day of the next month). 

Next, we compare ML models against DFM (see \autoref{tab:dmfml} in \autoref{apn:model_details}). Overall, DFM contributes to increasing prediction accuracy up to $25\%$ at $t+1$.\footnote{In this case, we use the DFM model with two factors. Including more factors does not improve results. Note: the DFM model's performance, in some cases, is similar to the OLS model.} However, in nowcasting GDP, RTS, and WTS at all three time horizons, the GBR, ENT, and feedforward ANN models---in many cases---perform better than DFM and other ML models considered. 
This is probably due to their ability to handle multiple predictors efficiently and capture sudden, large, and nonlinear interaction between the predictors and target variables during the COVID-19 crisis period. Overall, using payments data in the ML models, we observe up to a 12-30\% reduction in RMSE over DFM with the payments data.

Incorporating payments data in ML models provides downturn and recovery indications (during crisis periods) much better than the benchmark model in both in-sample and out-of-sample periods. We conjecture that this is due to the new and timely information provided by the payments data and ML model flexibility, allowing this data to provide better predictions during crisis periods. Visual comparisons of the best performing ML model against the benchmark model for in-sample and out-of-sample predictions are depicted in~\autoref{fig:testPredictions} in \autoref{apn:benchmarkvsml}. 

Next, we separately test our model's out-of-sample performance during a normal economic period (Jan 2019 to Feb 2020) and the COVID-19 period (Mar 2020 to Oct 2020) of the test sample (see~\autoref{tab:norvscovid} in \autoref{apn:normalvscrisisre}). 
We observe a higher gain using payments data during the time of crisis (up to 35\% RMSE reduction) compared to the normal period of the test sample (15--25\% reduction in RMSE) using payments data. These results suggest that payments data are useful during normal periods 
and its significance surges during periods of crisis, which substantially improves our model performance during those periods. 

Finally, we compare the GDP nowcasting performance of our model with the real-time vintages (first releases) and the latest vintages (see~\autoref{tab:realvslatest} in \autoref{apn:realtimevslatest}). Comparatively, the models using payments data perform better against the latest vintages. This makes sense, given that the latest vintages are more accurate (due to multiple revisions) compared to the real-time vintages. Therefore, we conclude that payments data are effective in providing timely estimates of key macro indicators. 

\subsection{Model Interpretation and Payments Data Contribution}
We now discuss the Shapley value-based interpretation of ML model predictions using the SHAP. We primarily focus on nowcasting GDP at time horizon $t+1$ using the tuned GBR model. 
However, we discuss a few key results for nowcasting RTS and WTS using a GBR, at the end of this section.  

\begin{figure}[htbp]
 \centering
 \includegraphics[width=0.65\textwidth]{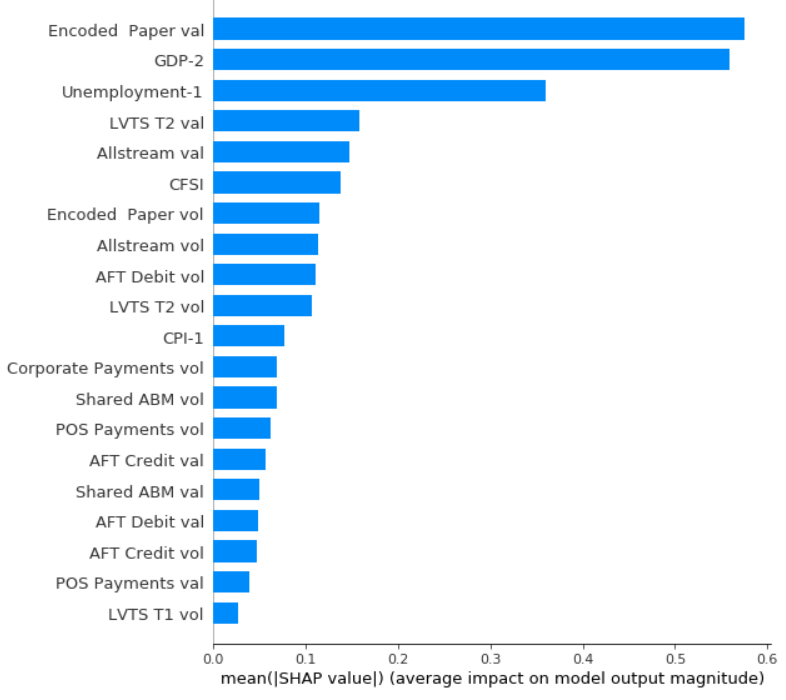}
 \caption{GDP: SHAP global feature importance measured as mean absolute Shapley values for each instance in the entire training sample (Mar 2005 to Dec 2020). The top 20 features are ranked from high (top) to low (bottom) based on average Shapley values.}
 \label{fig:summaryPlotAll}
\end{figure}

For demonstration, we use the entire sample (Mar 2005 to Dec 2020) for training. In~\autoref{fig:summaryPlotAll}, we plot the SHAP global feature importance obtained by averaging the absolute Shapley values for each predictor across the training set (in-sample).
This plot shows, on average, how much each feature influences the model prediction. These features are ranked according to their average influence (from high to low). For example, in the case of in-sample training data, both encoded paper value and GDP lag have a strong influence. However, the encoded paper stream is the strongest predictor (on average, it changes the GDP growth rate by  $0.6$ points). This is followed by the unemployment lag feature, LVTS-T2 value, and the sum of all the ACSS streams (Allstreams value).\footnote{For the tree-based models like GBR, we can also use impurity or permutation-based global feature importance~\cite{breiman2001random}. In our case, the permutation approach gives similar results as SHAP for the top three major predictors and matches eight out of the top ten highest contributors but slightly in a different order. Moreover, all three approaches rank the same three predictors in the top five list, and the Encode paper stream remains the most prominent feature in all approaches (see~\autoref{fig:featureImpComp}).}

\begin{figure}[htbp]
 \centering
 \includegraphics[width=0.65\textwidth]{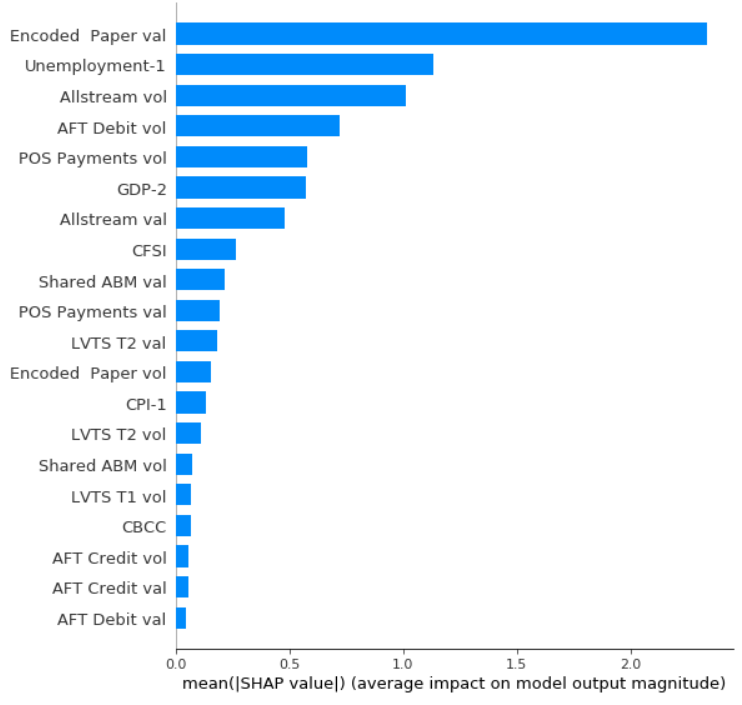} 
 \caption{GDP: SHAP global feature importance measured as mean absolute Shapley values of each instance in the training sample for the COVID-19 period (Mar 2020 to Dec 2020). The features are ranked from high (top) to low (bottom) based on average Shapley values.}
 \label{fig:summaryPlotCovid}
\end{figure}

In~\autoref{fig:summaryPlotCovid}, we show the global feature importance plot for the COVID-19 period with high negative growth rates (Mar to Dec 2020). During this crisis period, the influence of the encoded paper stream increases substantially along with the Allstream, AFT debit, and POS payments. 
GDP lag, the second most important feature for the entire training sample, loses its prediction power during the COVID-19 crisis period. A similar contribution from several payment streams is observed during the GFC period. These results suggest that, although lagged macro indicators influence GDP growth rates during normal periods, they do not add much value during crisis periods due to the delayed signal. During such periods, payments data become much more valuable and contribute well to macroeconomic prediction.

Next, using the SHAP \textquote{force} plots, we compute local feature importance, that is, the usefulness of each feature for a chosen sample point in the training set. Such insights could be important for nowcasting exercises because, during each step of the expanding window approach (i.e., when we advance by one month), the force plots could provide additional insights into each month's predictions by highlighting marginal contributions of individual predictors. 

\begin{figure}[htbp]
 \centering
 \includegraphics[width=0.99\textwidth]{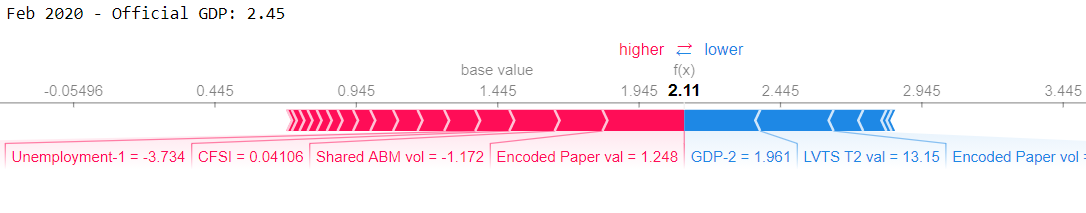} \includegraphics[width=0.99\textwidth]{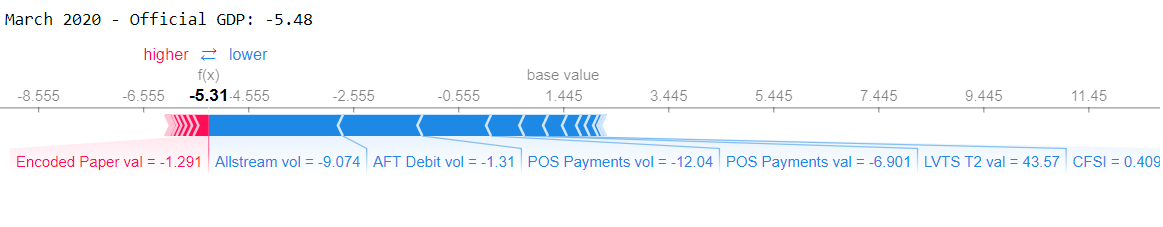}
 \includegraphics[width=0.99\textwidth]{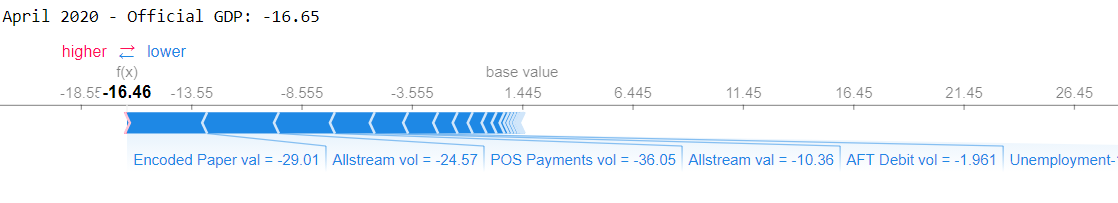} 
 \caption{GDP: SHAP force plots showing the feature contribution at each nowcasting instance during the onset of the pandemic, i.e., for Feb 2020 (top), Mar 2020 (middle), and Apr 2020 (bottom). The red arrows are positive Shapley values (contributing positively to GDP growth), and the blue arrows are negative Shapley values (contributing negatively to GDP growth). $f(x)$ is the model prediction at that instance, and the base value is the average of all predictions. Note: Values in red and blue are respective predictor values during that month; e.g., the Encode Paper value in Feb 2020 is 1.248.}
 \label{fig:forcePlots}
\end{figure}

For instance, in~\autoref{fig:forcePlots}, we plot the Shapley values as forces for Feb (top), Mar (middle), and Apr (bottom) 2020. Here, each Shapley value is an arrow that forces an increase (higher in red) or decrease (lower in blue) in the prediction from the baseline (the average of all predictions). The length of these arrows indicates the magnitude of the Shapley value for that feature.
These forces balance at the model prediction of that instance shown as $f(x)$. For Feb 2020, just before the pandemic began affecting Canada's economy, most payment predictors are positive (red) and pushing GDP growth higher. However, for Mar 2020 (the first month of the COVID-19 shock), most payment streams show a negative signal (blue). Similarly, for Apr 2020, all predictors show strong negative contributions, pushing the model prediction to the lowest value.  

\begin{figure}[htbp]
 \centering
 \includegraphics[width=0.99\textwidth]{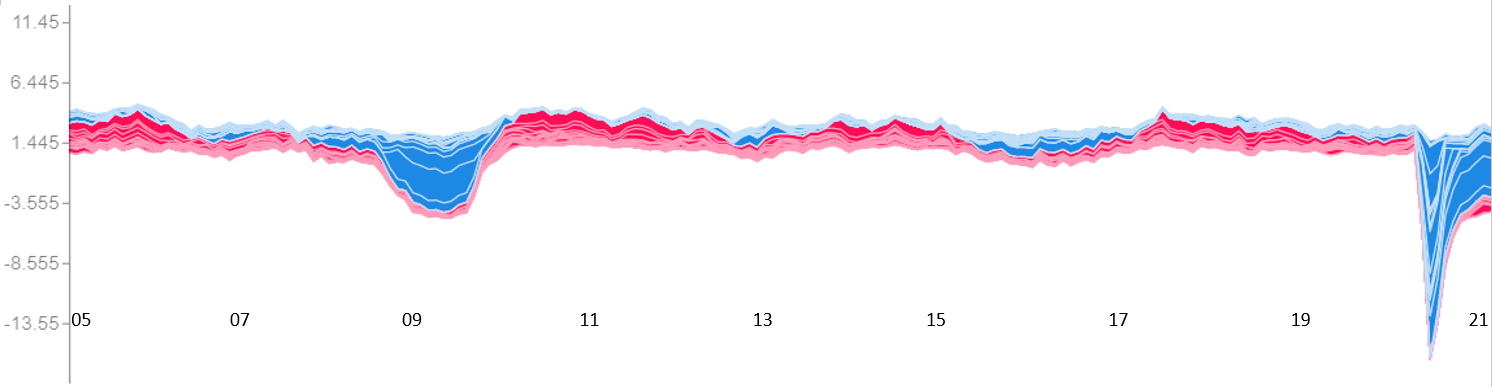} 
 \caption{GDP: Clustered force plots for each instance in the training sample, i.e., monthly instance from Mar 2005 to Dec 2020 positioned on the x-axis. Red clusters are positive Shapley values, i.e., the highest number of predictors contributing positively to GDP prediction, therefore pushing the prediction up, and blue clusters are negative Shapley values, i.e., the most predictors contributing negatively to GDP prediction, therefore bringing the prediction down (during the GFC and COVID-19 period). The line at the intersection of blue and red clusters is the actual model prediction.}
 \label{fig:clusterPlotTraining}
\end{figure}

\begin{figure}[htbp]
 \centering
 \includegraphics[width=0.49\textwidth]{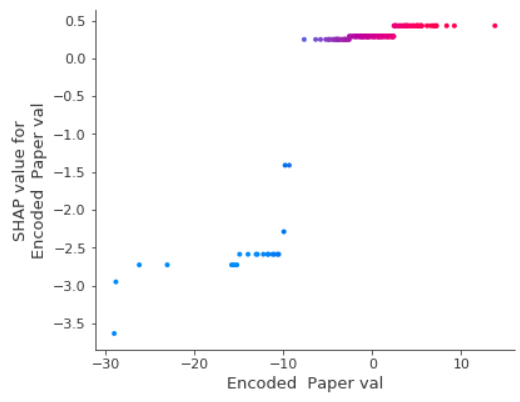} 
 \includegraphics[width=0.49\textwidth]{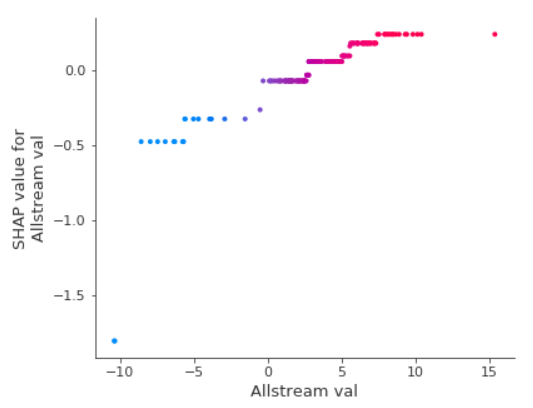} 
 \caption{Dependence plots show the Shapley value for each instance in the sample and the corresponding feature value. On the left, we show the dependence plot for the encoded paper (E) value, and on the right, we show the dependence plot for the ACSS Allstream (All) value.}
 \label{fig:dependenceplots}
\end{figure}

\begin{figure}[htbp]
 \centering
 \includegraphics[width=0.49\textwidth]{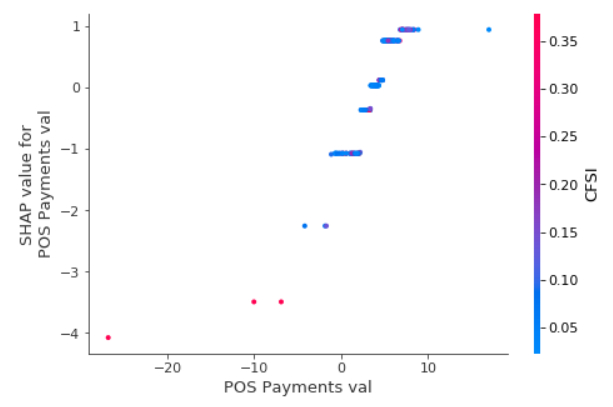} 
 \includegraphics[width=0.49\textwidth]{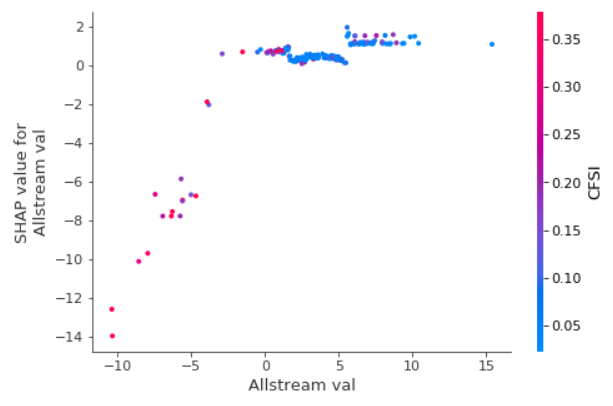} 
 \caption{Dependence plots show the Shapley value for each instance in the sample and corresponding predictor value. On the left we show a dependence plot of RTS for the POS payments value, and on the right we show the dependence plot of WTS for the ACSS Allstream value.} 
 \label{fig:dependenceplotsRTSWTS}
\end{figure}

\begin{figure}[htbp]
 \centering
 \includegraphics[width=0.49\textwidth]{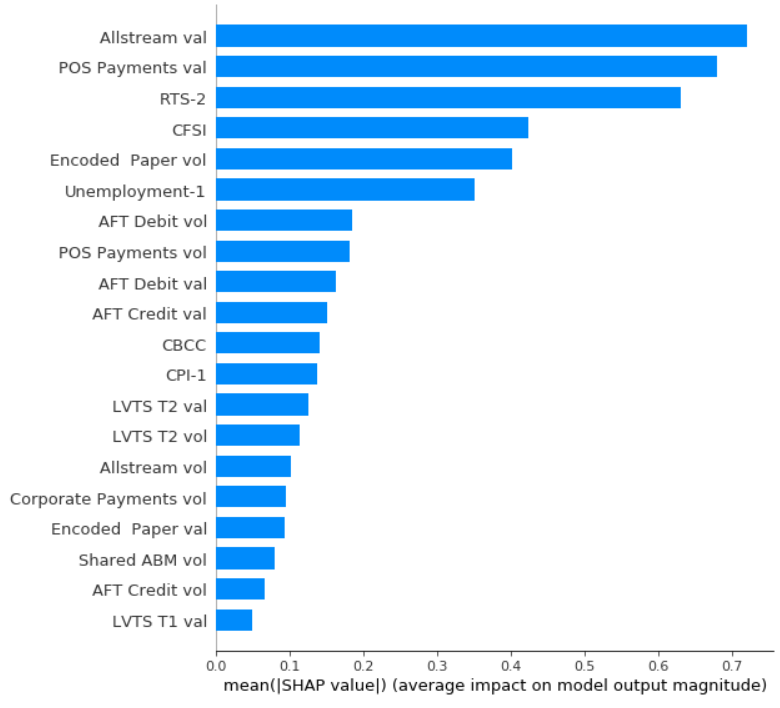}
 \includegraphics[width=0.49\textwidth]{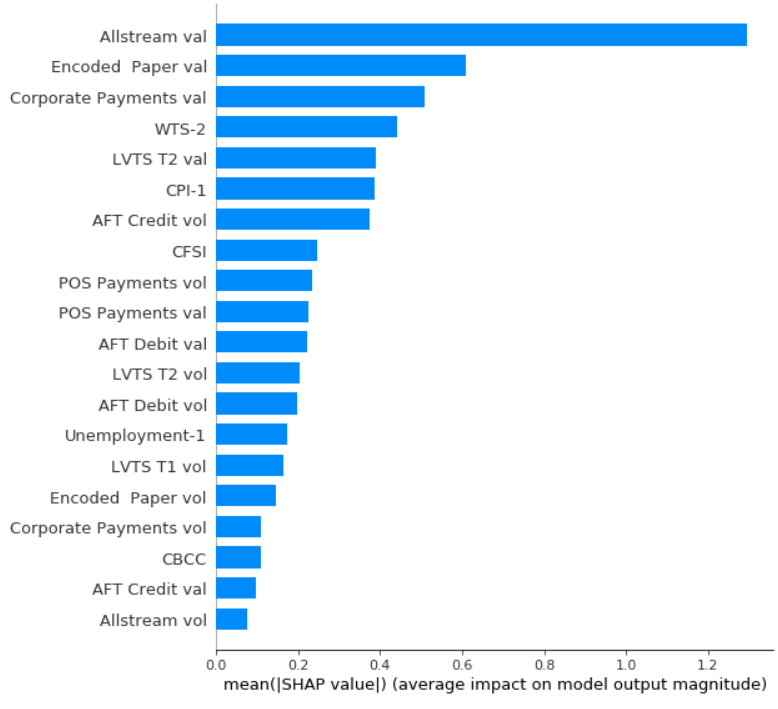}
 \caption{ (Top) Retail trade sales (RTS) and (bottom) wholesale trade sales (WTS). The SHAP global feature importance measured as the mean absolute Shapley value of each instance in the entire training sample (Mar 2005 to Dec 2020) at time horizon $t+1$ using the gradient boosting model.}
 \label{fig:summaryPlotRTS}
\end{figure}

\autoref{fig:clusterPlotTraining} shows the force plots for each instance in the entire training sample rotated and stacked together vertically. We observe red clusters of predictors with positive values (most predictors contributing positively to GDP prediction) during positive economic growth periods and blue clusters with negative signals during crisis periods, such as the GFC and COVID-19 shock. Such clustered signals could prove valuable in tracking crises in real time.

In \autoref{fig:dependenceplots}, we show dependence plots for encoded paper value (left) and Allstream value (right). These plots capture the relationship between the feature values on the x-axis and the corresponding Shapley values on the y-axis. Observe that the small and negative values of encoded paper growth rates provide higher contributions in Shapley values compared to the positive growth rates. However, both positive and negative growth rates of Allstreams value contribute similarly (or symmetrically).  
The encoded paper plot (left) suggests that the contribution of this stream, in terms of Shapley values, is small and linear during periods of normal growth. However, during periods of strong negative or positive growth ($>=|10|$), the contribution of this stream is asymmetrical and nonlinear.

Similar behavior is observed in nowcasting models for RTS and WTS using payments data and GBR model. In~\autoref{fig:dependenceplotsRTSWTS}, we show the dependence plots of RTS with POS payments value (left) and WTS with Allstream value (right). 
We also show how these payment streams are influenced by CFSI. These plots suggest that at high-stress levels, that is, at high values of CFSI (shown in red) and negative values of payment growth rates, the signal from these payment streams is strong and their contribution is high. However, for low levels of stress (blue) and high positive payment growth rates, the contributions from these streams are positive but small. This confirms the asymmetrical and potentially nonlinear relationship between these streams and the corresponding macro variables.

Finally, in~\autoref{fig:summaryPlotRTS}, we plot the SHAP global feature importance for the entire training set at time horizon $t+1$ for RTS (left) and WTS (right), respectively. These plots suggest, in the case of RTS, that the Allstream and POS payments values highly influence the model prediction. This makes sense, given that POS payments are commonly used for retail trade. In the case of WTS, the Allstream value has its strongest impact on the model prediction along with the encoded paper and corporate payments value streams, highlighting the importance of these streams in predicting WTS. These plots suggest that, in general, aggregated payments streams (Allstream and encoded paper) are crucial predictors in nowcasting GDP, RTS, and WTS.

\section{Conclusion}\label{sec:conclusion}

We use a set of comprehensive and timely payment systems data with ML models for macroeconomic nowcasting. The payments data provide information about the economy in real time and help reduce dependence on variables that are released with a significant delay. ML provides a set of advance econometric tools to effectively process various payment streams and capture the sudden, large, and nonlinear effects of a crisis. To improve the effectiveness of ML models for prediction, we use a Shapley value-based approach for model interpretability and a device specialized cross-validation strategy to avoid nowcasting model overfitting. 

Our results suggest that payments system data and ML models can lower nowcast errors significantly over benchmark models. We observe up to a 40\% reduction in RMSE over the linear benchmark in nowcasting GDP, RTS, and WTS. The most significant improvements are observed when we use the same month’s payments data as the target variable. Our nowcasting model out-of-sample performance is relatively higher during the COVID-19 period compared to the pre-COVID period. We also notice that the ML model performance changes slightly for different nowcasting cases. However, the gradient boosting regression model shows consistently good performance. The importance of payments data (especially the Encoded Paper stream) increases during crisis periods. Nonetheless, some payment streams strongly influence the model even during normal periods. 

We also demonstrate the usefulness of the Shapley value-based SHAP approach in gaining insights into ML model predictions at each nowcasting step and for the entire training sample. 
Further, we show how the dependence plots could help understand the relationship between the predictors' values and their influence on the target. Such insights could be valuable in macroeconomic monitoring and prediction, especially during crisis periods. 
Additionally, we find that the proposed cross-validation technique can help reduce overfitting and improve prediction accuracy in macroeconomic nowcasting models.
In conclusion, this paper substantiates the use of payments data and ML models for macroeconomic prediction and provides a set of econometric tools to overcome associated challenges to improve their effectiveness for policy use.

\newpage
\bibliographystyle{unsrt}  
\bibliography{references}

\newpage
\appendix
\section{Overview of ACSS and LVTS Payments Instruments}\label{apn:payments_det}
The historical list of payment streams processed through the ACSS payment system. Note: the first letter indicates the stream ID. This is followed by the stream label and a short description.
\begin{itemize}
\item  A: ABM adjustments - processes POS payment items used to correct errors from shared ABM network stream N.
\item B: Canada Savings Bonds - part of government items. Comprises bonds (series 32 and up and premium bonds) issued by the Government of Canada. Start date: April 2012.
\item C: AFT credit - processes direct deposit (DD) items such as payroll, account transfers, government social payments, business to consumer non‐payroll payments, etc.
\item D: AFT debit - pre-authorized debit (PAD) payments such as bills, mortgages, utility payments, membership dues, charitable donations, RRSP investments, etc.
\item E: Encoded paper - paper bills of exchange that include cheques, inter‐member debits, money orders, bank drafts, settlement vouchers, paper PAD, money orders, etc.
\item F: Paper-based remittances - used for paper bill payments, that is, MICR-encoded with a CCIN for credit to a business. It is similar to electronic bill payments (stream Y). 
\item G: Receiver General warrants - part of government payments items. Processes paper items payable by the Receiver General for Canada. Start date: April 2012.
\item H: Treasury bills and old-style bonds - part of government paper items. It processes certain Government of Canada paper payment items such as treasury bills, old-style Canada Savings Bonds, coupons, etc. Start date: April 2012.
\item I: Regional image captured payment (ICP) - processes items entered into the ACSS/USBE on a regional basis. Start date: Oct 2015. 
\item J: Online payments - processes electronic payments initiated using a debit card through an open network to purchase goods and services. Start date: June 2005.
\item K: Online payment refunds - processes credit payments used to credit a cardholder’s account in the case of refunds or returns of an online payment (stream J). Start date: June 2005.
\item L: Large-value paper - similar to stream E with value cap. Starting in Jan 2014, this stream merged into encoded paper stream E.
\item M: Government direct deposit - processes recurring social payments such as payroll, pension, child tax benefits, social security, and tax refunds. Start date: April 2012.
\item N: Shared ABM network - POS debit payments used to withdraw cash from a card-activated device.
\item O: ICP national - processes electronically imaged paper items that can be used to replace  physical paper items such as cheques, bank drafts, etc.
\item P: POS payments - processes payment items resulting from the POS purchase of goods or services using a debit card.
\item Q: POS return - processes credit payments used to credit a cardholder’s account in the case of refunds or returns of a POS payment (stream P).
\item S: ICP returns national - processes national image-captured payment returned items entered into the ACSS/USBE on a national basis. Start date: Oct 2015. 
\item U: Unqualified paper payments - processes paper-based bills of exchange that do not meet Canada payments association requirements for encoded paper classification.
\item X: Electronic data interchanges (EDI) payment - processes exchange of corporate‐to‐corporate payments such as purchase orders, invoices, and shipping notices.
\item Y: EDI remittances - processes remittances for electronic bill payments such as online bill and telephone bill payments.
\item Z: Computer rejects - processes encoded paper items whose identification and tracking information cannot be verified through automated processes.
\end{itemize}

The LVTS settles payments through two tranches, T1 and T2. Each tranche settles two types of payments: interbank and third-party funds transfers. The LVTS also includes transactions to and from the Bank of Canada~(See \cite{arjani2006primer} for more details.
\begin{itemize}
\item Foreign exchange payments and payments related to the settlement
of the Canadian-dollar leg of FX transactions undertaken in the continuous linked settlement (CLS) system.
\item Payments related to Canadian-dollar-denominated securities in the CDSX operated by clearing and depository services (CDS).
\item Payments related to the final settlement of the ACSS.
\item Large-value Government of Canada transactions (federal receipts and
disbursements) and transactions related to the settlement of the daily receiver.
\item The Bank of Canada’s large-value payments and those of its clients, which include Government of Canada, other central banks, and certain international organizations.
\end{itemize}

\section{Machine Learning Models}\label{apn:model_details}
In this section, we briefly discuss the ML models employed for nowcasting. For each model considered, there are many variations proposed in the literature. However, we have focused on the basic version of each model. Note that all models are implemented using the scikit-learn machine learning library~\cite{scikit-learn}. See~\autoref{apn:cross_validation} for more details on model training, tuning, and cross-validation procedures.

\subsection{Elastic Net Regularization}
Elastic net (ENT) is a regularized linear regression model. In ENT, the objective is similar to that of the OLS with the addition of $L_1$ and $L_2$ penalties. A regression model that uses only the $L_1$ penalty is called a Lasso regression, and a model that uses only the $L_2$ penalty is called a Ridge regression. In ENT, the combination of $L_1$ and $L_2$ penalties allows for learning a sparse model like Lasso, where only a few of the weights are non-zero. It also maintains the advantages of the Ridge regression, such as encouraging grouping effects, stabilizing regularization paths, and removing limitations on the number of selected variables~\cite{zou2005regularization,trevor2009elements}.

Consider a set $X = \{\bf{x}^1, \bf{x}^2, \dots, \bf{x}^M\}$ of $M$ attributes (independent variables) and a target $\bf{y}$ (dependent variable) and denote $\hat{\bf{y}}$ as the predicted target. 
With these specifications, in ENT, the objective function to minimize is
\begin{equation}\label{eq:enet}
  {\substack{\mathrm{min}\\ {\bf{w}}}}  \left\Vert {\bf{y}} - \hat{{\bf{y}}}(X,{\bf{w}}) \right\Vert^2_2 + \lambda_1 \left\Vert{\bf{w}}\right\Vert_1 + \lambda_2 \left\Vert{\bf{w}}\right\Vert^2_2,
\end{equation}
where ${\bf{w}}$ is a vector of unknown coefficients and $ \left\Vert . \right\Vert_*$ is $L_*$ norm. This procedure can be viewed as a penalized least squares method with the penalty factor $\lambda_1 \left\Vert{\bf{w}}\right\Vert_1 + \lambda_2 \left\Vert{\bf{w}}\right\Vert^2_2$. The ENT is particularly useful with a large set of predictors and correlated features. 
Note that we use the scikit-learn library for the implementation of ENT. We explore and tune the parameters $\lambda_1$ and $\lambda_2$ by controlling constant $\alpha$ that multiplies the penalty terms, mixing parameter $l1\_ratio$ and the maximum number of iterations. 
For example, the selected model for GDP nowcasting At $t+1$: $alpha$ is 0.001, $l1\_ratio=0.5$.
For other parameters, we use the default values~\cite{scikit-learn}. 

\subsection{Support Vector Regression}
Support vector regression (SVR) is another model useful when there are problems with multiple predictors. It uses a different objective function than the ENT. The SVR is based on support vector machines where the task is to find a hyperplane that separates the entire training dataset into, for example, two groups by using a small subset of training points called support vectors.
In SVR the goal is to find a function, for instance, the linear function $f({\bf{x}}_i) = {\bf{w}}^{T} {\bf{x}}_i + b$ (where $b$ is a bias and $i=1, 2, \dots N$) that has at most $\varepsilon$ deviation from the actual ${\bf{y}}$ for all the training data.
Therefore, the objective function to minimize is
\begin{equation}\label{eq:svm}
   \frac{1}{2} \left\Vert {\bf{w}} \right\Vert^2_2 + C \sum_{i=1}^N |{\bf{y}}_i - f({\bf{x}}_i)|_{\varepsilon},
\end{equation}
subject to
\begin{eqnarray}\label{eq:svm_c}
  {\bf{y}}_i - f({\bf{x}}_i) \leq \varepsilon  \\
  f({\bf{x}}_i) - {\bf{y}}_i \leq \varepsilon,
\end{eqnarray}
where $N$ is the number of training samples and $C$ is a regularization parameter constant~\cite{smola2004tutorial}.
A different type of kernel function (linear, polynomial, sigmoid, etc.) can be specified for the decision function. Therefore, it is versatile. For further details of SVM theory and formulation, see \cite{smola2004tutorial,trevor2009elements}.

Note: we use scikit-learn library for implementing SVR. We explore and tune the following hyperparameters: kernel type, degree of the polynomial kernel function, and regularization parameter constants $C$ and $\epsilon$. 
For example, the selected model for GDP nowcasting At $t+1$: kernel is $rbf$, degree = $2$, $C$ = 3, and $\epsilon=0.3$.
We use the default values for other parameters. 

\subsection{Random Forest}

\begin{figure}[htbp]
 \centering
 \includegraphics[width=0.62\textwidth]{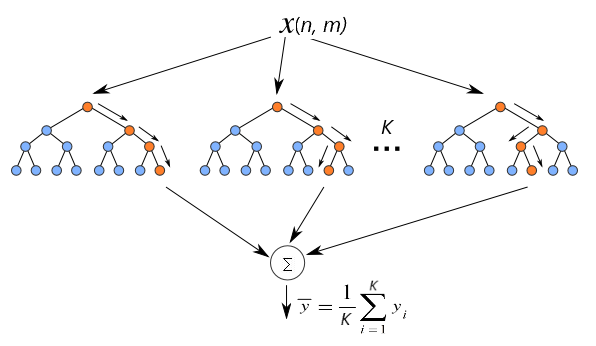}
 \caption{Random forest with $K$ trees using $n$ samples and $m$ features for each tree.}
 \label{fig:rf_basic}
\end{figure}

Another popular approach is random forest regression (RFR). It is a decision tree (DT)-based ensemble learning method built using a forest of many regression trees. It is a non-parametric method and hence approaches the multicollinearity problem slightly differently from parametric approaches such as OLS and ENT.
In RFR, each DT is independently built from a bootstrapped subset of the training set. Each bootstrap sample can randomly select a subset of features from the available set or the full features set. The final prediction is performed by averaging the predictions of all regression trees.
The procedure is visually depicted in \autoref{fig:rf_basic}.
The two levels of randomness (i.e., a random subset of the sample and the features) incorporated to build the DT can help to reduce variance in the predictions.
RFR has been shown to handle highly nonlinear interactions between multiple predictors and a target variable~\cite{breiman2001random,liaw2002classification}.

Note: we use scikit-learn library for the implementation of the RFR regression. We explore and tune the following hyperparameters: the number of trees in the forest ${n\_estimators}$, the maximum depth of the tree $max\_depth$, and the minimum number of samples required to split an internal node $min\_samples\_split$. 
For example, the selected model for GDP nowcasting At $t+1$: ${n\_estimators}$ is 400, $max\_depth$ is 4 and $min\_samples\_split$ is 2. We use the default values for other parameters. 

\subsection{Gradient Boosting}

Similar to RFR, gradient boosting regression (GBR) is a DT-based non-parametric ensemble learning approach.
It is a general technique of boosting in which a sequence of weak learners (e.g., small DTs) are built on a repeatedly modified version of the training set.
The data modification at each boosting interaction consists of applying weights to each of the training samples, and for successive iterations, the sample weights are modified. Basically, the next learner is fit on the residual of the previous learner~\cite{friedman2001greedy,friedman2001elements}.

GBR trees are additive models whose prediction $\hat{\bf{y}}$ for a given input $X$ for each instance $i$ can be written as
\begin{eqnarray}\label{eq:gb1}
\hat{\bf{y}}_i = H_p(X_i) = \sum_1^{p} h_p(X_i),
\end{eqnarray}
where $h_p$ are weak learners, for example, decision trees~\cite{friedman2001elements} and $p$ is the number of learners. 
The model $H_P(X)$ is built as
\begin{eqnarray}\label{eq:gb2}
H_p(X) = H_{p-1}(X) + \gamma h_p(X),
\end{eqnarray}
where $\gamma$ is the learning rate used to regularize the contribution of each new weak learner, and the newly added weak learner $h_p$ (tree) is used in order to minimize a sum of losses $L_p$:
\begin{equation}\label{eq:gb3}
 h_p = {\substack{\mathrm{arg \ min}\\ {\bf{p}}}} L_p. 
\end{equation}

Both RFR and GBR techniques are interpretable to a certain extent because these models use DTs as their base learners. The DTs perform feature selection from the set provided by selecting appropriate split points. This information can be used to measure the importance of each feature~(see \cite{scikit-learn} for additional details).
Note: we use scikit-learn library for implementation of GBR. We explore and tune the following hyperparameters: the number of trees in the forest $n\_estimators$, the maximum depth of the tree $max\_depth$, and the learning rate, which helps shrink the contribution of each tree. For example, the selected model for GDP nowcasting At $t+1$: ${n\_estimators}$ is 1000, $max\_depth$ is 1, and $learning\_rate$ is 0.1. We use the default values for other parameters.

\subsection{Feed-Forward Artificial Neural Network}
\begin{figure}[htbp]
 \centering
 \includegraphics[width=0.53\textwidth]{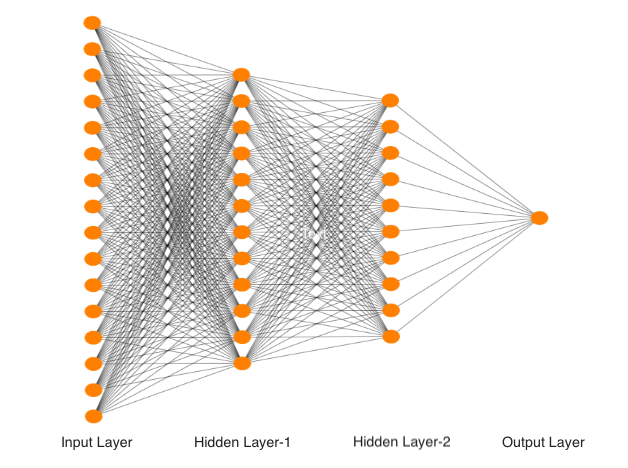}
 \caption{Schematic of densely connected feed-forward neural network with two hidden layers.}
 \label{fig:ann-basics}
\end{figure}
A feed-forward artificial neural network (ANN) with hidden layers is multiple layers of artificial neurons sandwiched between input and output layers, as depicted in~\autoref{fig:ann-basics}.
In a feed-forward ANN, the data always moves forward through the network layers. It starts in the input layer, for instance, each input feature instance ${\bf{x}}_i$ is multiplied by its corresponding layer's  weight ${\bf{w}}$. Then, the weighted sum of these inputs ${\bf{w}}^T {\bf{x}}_i + b$ (where $b$ is a bias) is passed through a nonlinear activation function $\sigma$, resulting in a nonlinear function of the inputs $\sigma({\bf{w}}^T {\bf{x}}_i + b)$.
Then the outputs are sent to the next layer. This process continues until the last layer.
Once we get the final output from the network, denoted as $\hat{{\bf{y}}}$, we measure how good that output is compared to the actual value of the target ${\bf{y}}$. This is done by using an objective function, for example, mean squared error. Given these results, we go back and iteratively adjust the weights and biases of the network to optimize the objective function. For further details on the activation function and optimization procedure, see~\cite{bengio2009learning,goodfellow2016deep}.

The greater the number of layers, the deeper the network. Therefore, it is generally referred to as the deep neural network (DNN).
The multilayer architectures enable a combination of features from lower layers, potentially modeling complex data with fewer units. Therefore, the DNN can be used to model complex nonlinear relationships between the input and output. However, DNN requires tuning a large number of hyperparameters as the number of hidden layers grows. Therefore, generally, it needs a large training dataset to achieve a good performance.

Note: we use scikit-learn's multi-layer perception (\emph{MLPRegressor}), and we explore and tune the following hyperparameters: The number of neurons in the hidden layers $hidden\_layer\_sizes$, the activation function for the hidden layer $activation$, and the learning rate schedule for weight updates. 
For example, the selected model for GDP nowcasting At $t+1$: $hidden\_layer\_sizes$ is 3, $activation$ is $relu$, and $learning\_rate$ is 0.05. We use the default values for other parameters.

\subsection{ML Models Performance Caparison with DFM}

Here we compare ML models' performance against DFM with the payments data (main case). Similar to the model employed in~\cite{chernis2017dynamic}, we use the DFM model with two factors (including additional factors does not improve model performance) and one lag in the VAR driving the dynamics of those factors. Idiosyncratic components are assumed to follow an AR(1) process. In nowcasting GDP, RTS, and WTS, the GBR, ENT, and feedforward ANN models---in many cases---perform better than DFM and other ML models considered. Overall, using payments data in the ML models, we observe up to a 12-30\% reduction in RMSE over DFM with the payments data.

\vspace{2mm}
\rowcolors{2}{gray!10}{white}
\begin{table}[htbp]
  \begin{center}
  \begin{threeparttable}
    \caption{Out-of-sample RMSE comparisons of DFM with ML models for seasonally adjusted YOY growth rate of macro variables at the horizons $t+1$ (top panel), and $t+2$ (bottom panel) for the main case\tnote{a} }
    \label{tab:dmfml}
      \begin{tabular}{c c c c c c c c}
      \toprule
      \addlinespace
      \textbf{Target}\tnote{b}  \ & \ \textbf{DFM}\tnote{c} \ & \ \textbf{ENT}\tnote{d}  \ & \ \textbf{SVR}\tnote{d} \ & \ \textbf{RFR}\tnote{d} \ & \ \textbf{GBR}\tnote{d} \ & \ \textbf{ANN}\tnote{d} \ & \ \textbf{\% Reduction}\tnote{e} \\
      \addlinespace
      \bottomrule
         GDP & 1.00 & \textbf{0.96} & 1.41 & 1.11 & \textbf{\underline{0.81}}\tnote{f} & \textbf{0.82} & 19 \\
         RTS & 1.00 & \textbf{0.89} & 1.27 & 1.07 & \textbf{\underline{0.85}} & 1.02 & 15 \\
         WTS & 1.00 & \textbf{0.96} & 1.14 & \textbf{0.82} & \textbf{0.69} & \textbf{\underline{0.51}} & 31 \\
      \bottomrule
      \end{tabular}
          \begin{tabular}{c c c c c c c c}
      \toprule
      \addlinespace
      \textbf{Target} \ & \ \textbf{DFM} \ & \ \textbf{ENT} \ & \ \textbf{SVR} \ & \ \textbf{RFR} \ & \ \textbf{GBR} \ & \ \textbf{ANN} \ & \ \textbf{\% Reduction} \\
      \addlinespace
      \bottomrule
         GDP & 1.00 & \textbf{0.87} & 1.62 & 1.14 & \textbf{\underline{0.82}} & \textbf{0.85} & 18 \\
         RTS & 1.00 & \textbf{\underline{0.87}} & 1.36 & 1.15 & \textbf{0.90} & \textbf{0.97} & 11 \\
         WTS & 1.00 & \textbf{0.89} & 1.19 & \textbf{0.91} & \textbf{0.81}  & \textbf{\underline{0.70}} & 19 \\
      \bottomrule
      \end{tabular}
    \begin{tablenotes}\footnotesize
    \item [a] In-sample training period, Mar 2005 to Dec 2018, ($p=166$) and out-of-sample testing period, Jan 2019 to Dec 2020, ($p=24$). All RMSEs are normalized with respect to DFM. The performance gain using ML models for time horizon $t$ are much smaller, however, GBR model performed better compred to other ML models. 
    \item [b] RTS-retail trade sales, WTS-wholesale trade sales. Note: we use the latest available values of targets for these exercises.
    \item [c] For DFM, we use payments data along with the predictors in the benchmark case. We use the DFM model with two factors and one lag in the VAR driving the dynamics of those factors. Idiosyncratic components are assumed to follow an AR(1) process.
    \item [d] We use elastic net (ENT), support vector regression (SVR), random forest regression (RFR), gradient boosting regression (GBR), and ANN. For these ML models, we select the model parameters and number of payment predictors based on target variables using the cross-validation procedure outlined in~\autoref{sec:methods}. Further details on these models are provided in~\autoref{apn:model_details}. Model selection and cross-validation procedures are detailed in~\autoref{apn:cross_validation} and~\ref{apn:selectBestFeatures}.
    \item [e] Percentage reduction in RMSE over DFM for GBR model. 
    \item [e] The models with out-of-sample prediction RMSE less than DFM ($<1$) are highlighted (bold) and the best model is also underlined.
    \end{tablenotes}
    \end{threeparttable}
  \end{center}
\end{table}

\section{Model Parameter Selection and Cross-Validation}\label{apn:cross_validation}
\begin{figure}[htbp]
 \centering
 \includegraphics[width=0.99\textwidth]{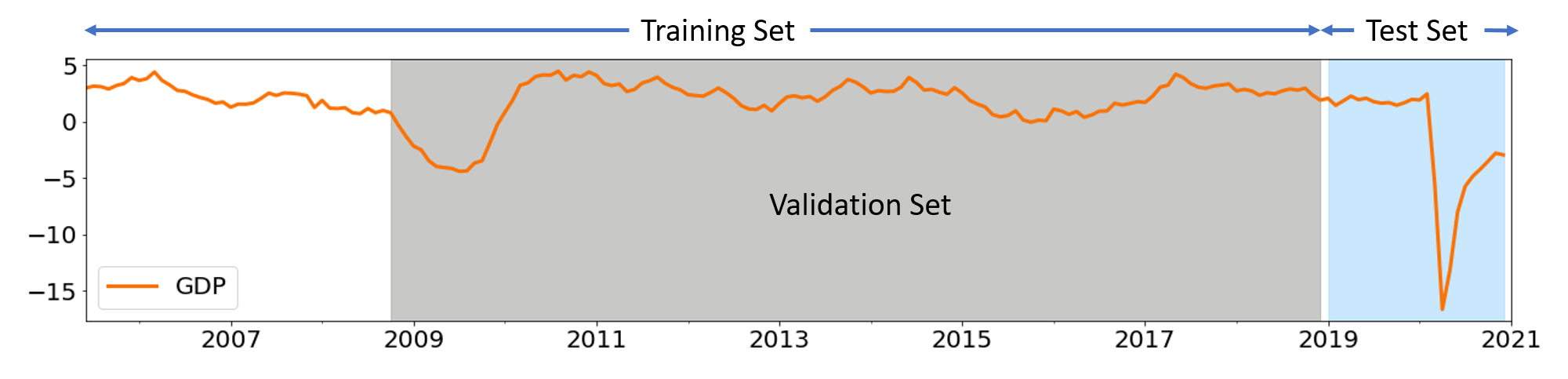}
  \caption{Schematic of data splits for cross-validation. First, the dataset is divided into a training set with a validation subset sampled from the highlighted gray area and a test set (highlighted in blue). The orange line shows the GDP growth rate.}
 \label{fig:gdp_split}
\end{figure}

The hyperparameter tuning and cross-validation of each ML model employed in this paper are performed using the randomized expanding window approach with $k$-folds as follows:
\begin{enumerate}
  \item Split the original dataset into a training set and test set (\autoref{fig:gdp_split}). In our case, the training set is Mar 2005 to Dec 2018, and the test set is Jan 2019 to Dec 2020 (highlighted in blue). 
  \item Specify the hyperparameters to tune and select the range for each parameter. See~\autoref{apn:model_details} for individual model parameters selected for tuning.
  \item Select two dates in the training set that define the validation superset (highlighted in gray in~\autoref{fig:gdp_split}). To include the global financial crisis, we choose those dates between Oct 2008 and Dec 2018.
  \item Next, for each fold in the cross-validation, we randomly sample 24 points (it is the same as the test set) from the validation superset as the validation subset (see~\autoref{fig:cv_expanding_window} for an example). 
  \item Using the selected parameters grid and validation subset, 
  we do the following: \\
  (a) For each iteration in the expanding window over the validation subset, select a data point from that subset as the out-of-sample test point and use all the data points up to that point for training (see \autoref{fig:TimeSeriesSplit_validation} where red dots are test points and blue dots are training points). \\
  (b) Fit the model on the selected training sample. \\
  (c) Using the trained model, predict for the selected sample point in the validation subset. \\
  (d) Repeat steps a, b, and c for each point in the validation subset. \\
  (e) After finishing iterating the chosen validation subset, compute the validation RMSE.
\item Repeat steps 4 and 5 $k$-times (typically $k$ is between 5 to 10), each time using a new validation subset. 
\item Compute the average validation RMSE over the $k$-folds. 
\item Select the parameters for which the average validation RMSE is smallest.
\item Use the tuned model to get the RMSE for the testing set by reusing the standard expanding window approach, as illustrated in \autoref{fig:TimeSeriesSplit_validation}.
\end{enumerate}

\begin{figure}[htbp]
 \centering
 \includegraphics[width=0.6\textwidth]{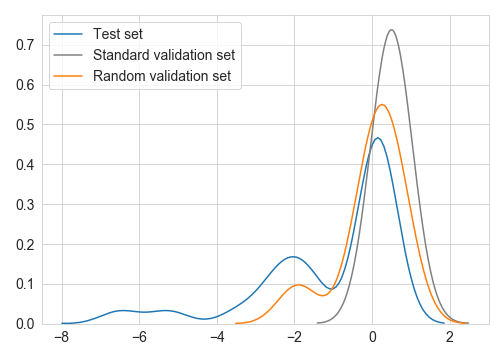}
  \caption{GDP: Distribution of the standardized test set and a typical validation set for standard $k$-fold expanding window approach (standard validation set) and random expanding window approach proposed here (random validation set). The distribution of the proposed random validation set remains similar across all $k$-folds; however, the distribution of the standard validation set could change based on the sample period.}
 \label{fig:gdp_cv_sample_distributions}
\end{figure}

In~\autoref{fig:gdp_cv_sample_distributions}, we present standardized distribution of the target variables (GDP growth rate) for the out-of-sample testing period (Jan 2019 to Dec 2020). In the same figure, we plot the distribution for a validation sample of the standard expanding window approach and the proposed randomized expanding window approach for cross-validation. The distribution of each of the randomized validation sets typically contains a few sample points from 2008 GFC; therefore, it is skewed towards the left---similar to the test set; however, this is not true for each standard validation set, except for the first validation set where we have the entire 2008 GFC period (see~\autoref{fig:cv_expanding_window}).   

\section{Feature Selection}\label{apn:selectBestFeatures}
To select the $k$-best predictors from the set of available attributes, we employ the {\it{SelectKBest}} method from scikit-learn~\cite{scikit-learn}. This method removes all but the $k$ highest-scoring features using univariate linear regression tests. It is a linear model for testing the individual effect of each of many regressors.
To select the $k$-best variables, we employ the following steps:
First, the correlation between each predictor and the target is computed. 
Next, the computed correlations are converted to $F$-scores (using the $F$-test), then to $p$-values.
Finally, these $F$-scores with $p$-values are used to select the $k$ highest-scoring features.

\begin{figure}[htbp]
 \centering
 \includegraphics[width=0.99\textwidth]{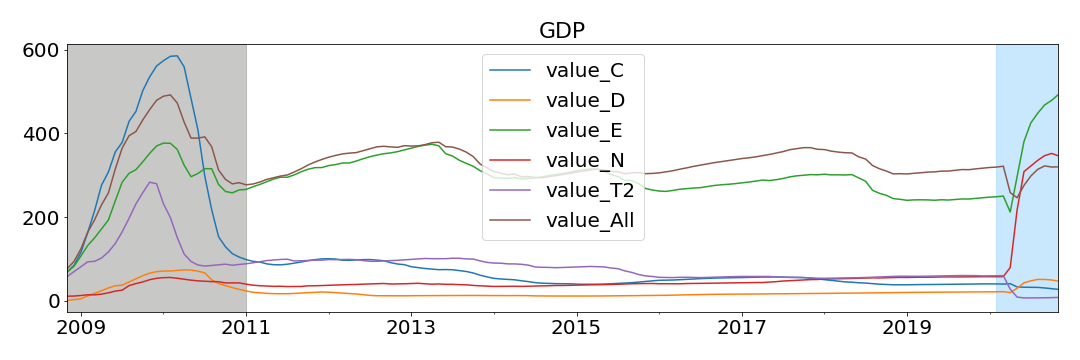}
 \includegraphics[width=0.99\textwidth]{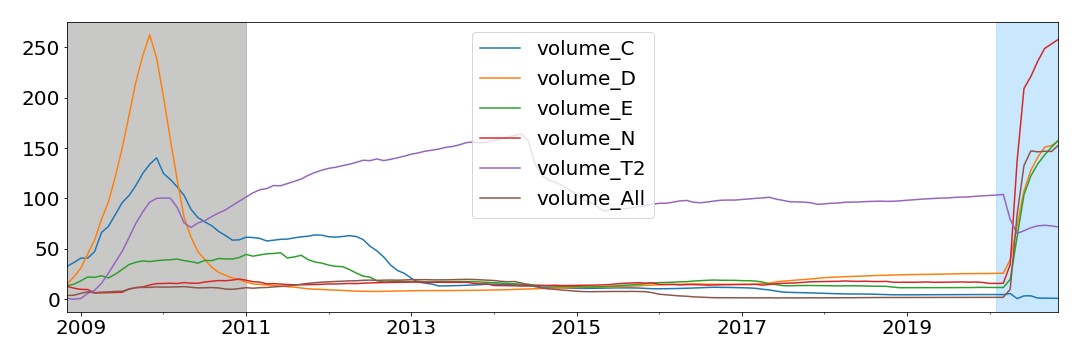}
  \caption{The $F$-score of a few selected payments streams (values-top, volumes-bottom) for GDP nowcasting. Higher scores mean a high prediction value. These plots are obtained after each training session of the expanding window approach, ranging from Oct 2008 to Dec 2020. The 2008 GFC period is highlighted in gray; blue shows the COVID-19 period.}
 \label{fig:gdp-une_score}
\end{figure}

In \autoref{fig:gdp-une_score}, we plot the scores of a few of the selected {\it{value}} streams (top) and {\it{volume}} streams (bottom) for GDP over the expanding window for the period ranging from Oct 2008 to Dec 2020.
The prediction scores for most of the value and volume streams are high during the GFC. The scores are steady and low during normal times (2011--2019) except for the encoded paper value (E), Allstream value (All), and LVTS-T2 volume (T2), for which scores remain high. During the COVID-19 crisis (Mar to Dec 2020), however, we see opposite behaviour in the prediction scores of a few streams. For example, AFT credit (C) and LVTS-T2 value streams have strong prediction scores during the GFC. However, their scores are weak during the COVID-19 period. Similarly, the ABM stream (both value and volume) has low scores during the GFC, but the scores are high during the COVID-19 period. 

\section{The Shapley Values and SHAP for Model Interpretation}\label{apn:shapvalue}
The Shapley values is a method from coalitional game theory that provides a way to fairly distribute the \emph{payout} among the \emph{players} by computing the average marginal contribution of each player across all possible coalitions~\cite{shapley1953value,osborne1994gametheory}.

For a coalitional game, $(N, v)$, where $N$ is a finite set of players indexed by $i$ and $v$ is the utility function or payoff function, the Shapley value can be obtained by this theorem, which satisfies the symmetry, dummy, and additivity axioms~\cite{osborne1994gametheory}: 
\begin{eqnarray*}
\phi_i(N, v) = \underbrace{\frac{1}{N!} \sum_{S \subseteq N \backslash \{i\}} }_{\textrm{average over all} \ S} \underbrace{|S|! \ \Big(|N| - |S| - 1\Big)!}_{\textrm{possible coalitions}} \ \ \underbrace{\bigg[ v(S \cup \{i\}) - v(S) \bigg] }_{\textrm{marginal value}}. 
\end{eqnarray*}

At a high level, the above equation can be split into three parts. The last part of the equation (the marginal value) gives the marginal contribution of an individual player $i$, when added to the coalition $S$ that does not have $i$. The middle part shows how to compute different possible ways in which we could have formed the coalitions. Then, we take an average of possible ways that we could have done the marginal value calculation. 

The SHAP proposed by~\cite{lundberg2020local2global} uses the Shapley values to explain the model predictions in terms of the marginal contribution of each predictor. The SHAP specifies the explanation of model $\mathcal{F}$ as a linear model of coalitions: 
\begin{eqnarray}
    \mathcal{F}(S) = \phi_0 + \sum_{i=1}^M \phi_i S_i,
\end{eqnarray}
where $S \in \{0, 1\}^M$ is the coalition vector with maximum $M$ coalitions and $\phi_i$ the Shapley value for $i^{th}$ player. In $S$ the entry 1 means the corresponding player is present and 0 means the player is absent. 

\begin{figure}[htbp]
 \centering
 \includegraphics[width=0.75\textwidth]{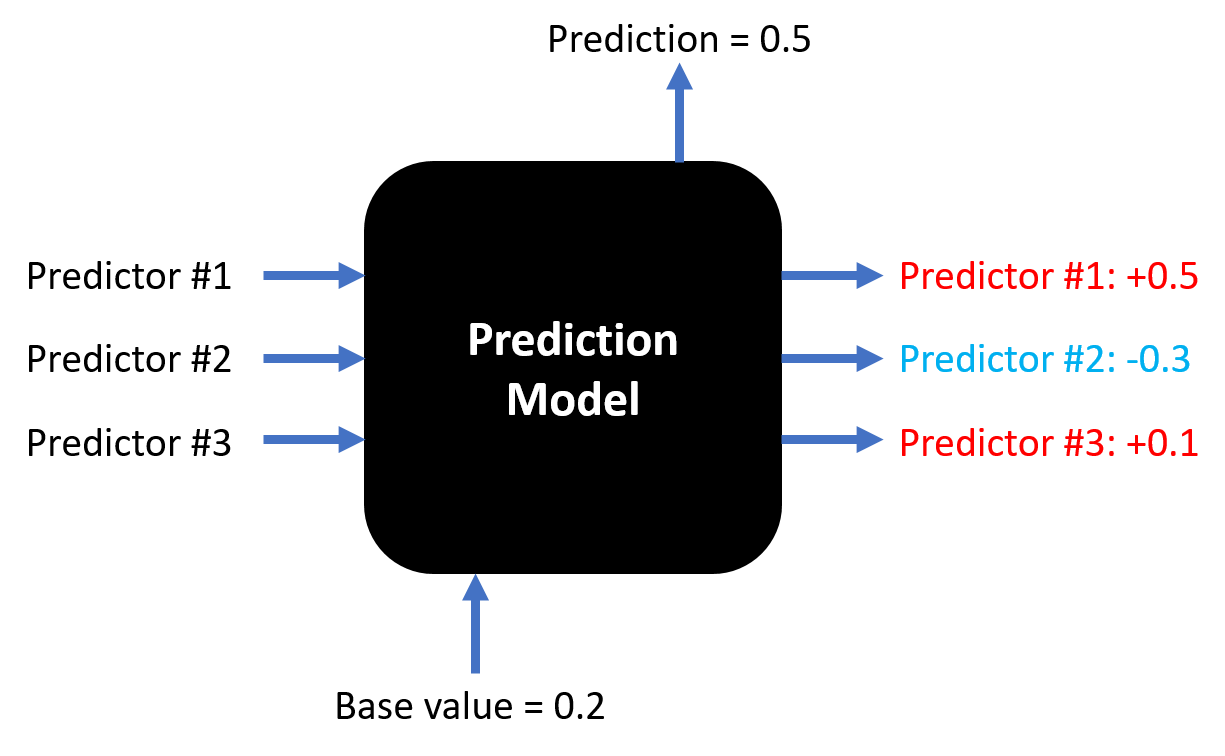}
  \caption{The SHAP explainer provides the marginal contribution of each predictor.}
 \label{fig:shapart}
\end{figure}

To illustrate, consider nowcasting is a {\emph{game}}. Then the Shapley values can be used to fairly distribute the \emph{payout} (= the prediction) among the  \emph{players} (= the predictors). Note: for the computation of the Shapley values in the SHAP, the zero means the corresponding predictor is absent. In that case, the absent predictors' value is replaced by a random value from its sample~\cite{lundberg2020local2global,molnar2020interpretable}. The procedure is further illustrated as follows:

\begin{enumerate}
	\item  Consider a nowcasting problem with three predictors (\autoref{fig:shapart}) in a prediction model (it could be any model) to predict a target (for instance, monthly GDP growth).
	\item  The average prediction of the model, that is, the base value is 0.2, and for the current instance (for example, month $t$), our model predicts GDP growth 0.5. 
	\item  By computing the Shapley values for all possible coalitions among three predictors, we can explain the difference between actual prediction (0.5) at current month $t$ and the base value (0.2) in terms of each predictor's contribution. 
	\item  In the current example, predictor 1 increases the growth rate by 0.5 percentage points, predictor 2 pushes it down by 0.3 points, and predictor 3 contributes +0.1 points. Thus, together these three predictors increase the prediction by +0.3 points from the average predictions of the entire sample of 0.2, leading to the final prediction of 0.5 growth. 
\end{enumerate}

The SHAP values tell us which predictor contributes the most in the current instance of the prediction, that is, a local interpretation.  Similarly, by using the Shapley values for each instance in the sample, we can get the average contribution of each predictor for that sample. That would give us a global interpretation of the model in terms of its feature importance. However, it is important to remember that these are only for the chosen model, and they do not explain the causality.

The SHAP package developed by~\cite{lundbergNips2017unified,lundberg2020local2global} provides various tools to visualize the Shapley values computed for various ML models commonly used for predictions. For instance, the feature importance plots and summary plots (\autoref{fig:summaryPlotAll} and \ref{fig:summaryPlotCovid}) are useful for global model interpretations. The force plots or clustered force plots (\autoref{fig:forcePlots} and \ref{fig:clusterPlotTraining}) are useful for local interpretation, that is, at each instance of prediction. 
Also, the dependence plots (\autoref{fig:dependenceplots}) could be valuable for understanding the relationships between given predictors and the targets. 

The SHAP, although a powerful model-agnostic ad-hoc tool developed based on theoretical foundations for model interpretability, has some shortcomings, and it should be used with caution~\cite{molnar2020interpretable,slack2020fooling}. For example, the KernelSHAP is computationally intensive and could be very slow for problems with many predictors. However, for macroeconomic predictions models, we have comparatively fever predictors (20--50) and fewer instances (a few hundred data points). Therefore, it is not much of an issue in such applications. Another issue with KernelSHAP is that it is sensitive to colinearity in the predictors. The TreeSHAP approach developed in~\cite{lundberg2020local2global} overcomes some of those challenges to a certain extent~\cite{molnar2020interpretable}. Furthermore, as shown by~\cite{slack2020fooling}, it is possible to misuse such ad hoc tools to hide model biases. However, it is not much of a concern for the macroeconomic prediction models we deal with in this paper. Additionally, the authors conclude that the SHAP is less prone to such problems than several other interpretation tools. 

\subsection{Global Feature Importance Comparison}\label{apn:shapvsimputance}
\begin{figure}[htbp]
 \centering
 \includegraphics[width=0.99\textwidth]{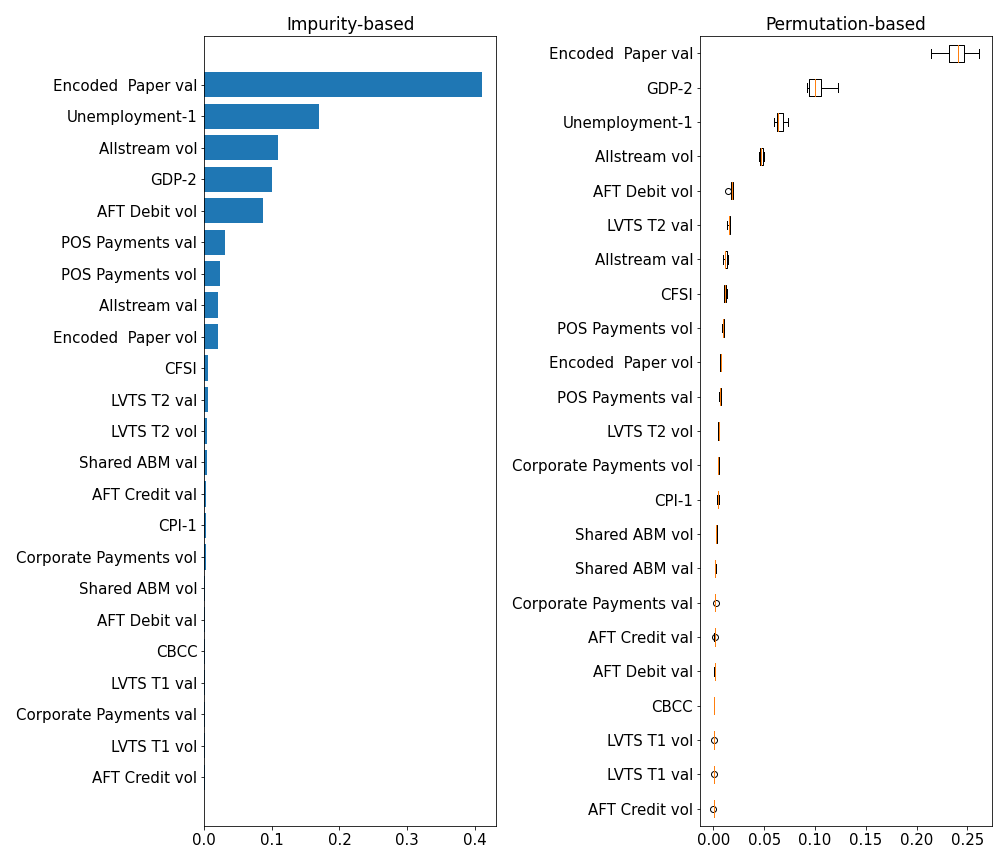}
 \caption{GDP: Global feature importance for the entire training sample (Mar 2005 to Dec 2020) at time horizon $t+1$ using the gradient boosting model. (Left) impurity-based feature importance and (right) permutation-based feature importance.}
 \label{fig:featureImpComp}
\end{figure}

We can also use impurity, or permutation-based global feature importance approaches for tree-based models like GBR and RFR. 
Amongst the two the permutation-based approach is shown to be more useful for nonlinear models~\cite{breiman1996bagging,molnar2020interpretable}. 
In \autoref{fig:featureImpComp}, we compare the feature importance of the gradient boosting model trained on the entire training sample (Mar 2005 to Dec 2020) at time horizon $t+1$.
The permutation-based approach is similar to SHAP for the top three major contributors and matches eight out of the top ten highest contributors but slightly in a different order. Moreover, all three approaches rank the same three predictors in the top five list, and the Encode paper stream remains the most prominent predictor in all three approaches.

\section{Nowcasting Performance for Benchmark and ML Models}\label{apn:benchmarkvsml}
\begin{figure}[htbp]
 \centering
 \includegraphics[width=0.99\textwidth]{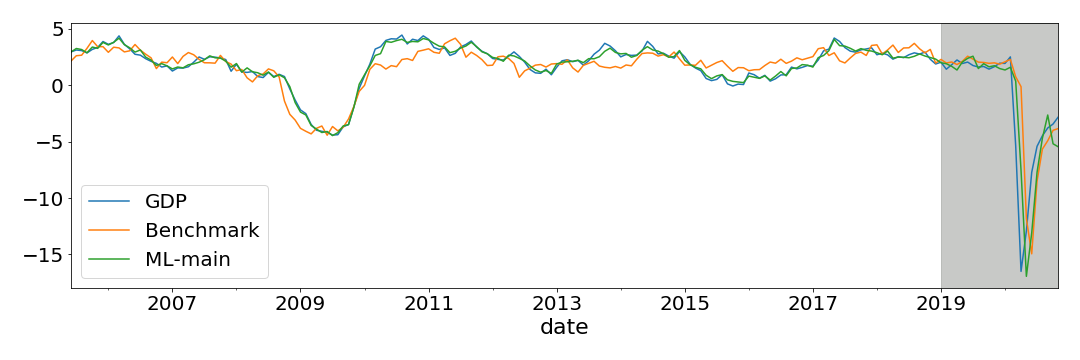}
 \includegraphics[width=0.99\textwidth]{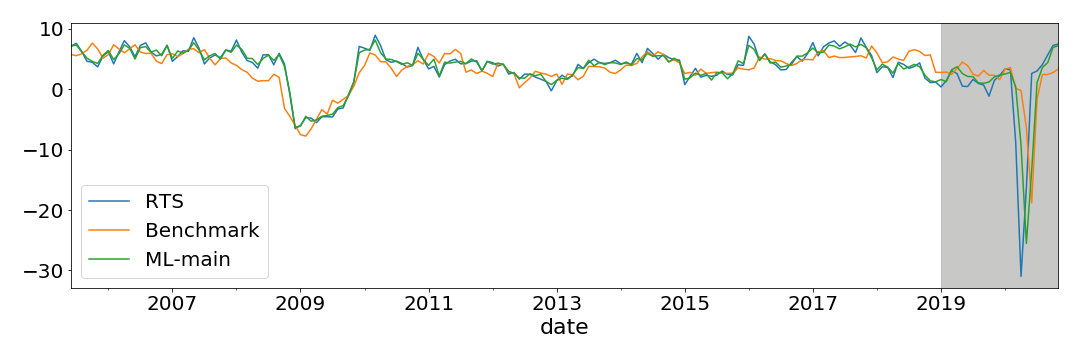} 
 \includegraphics[width=0.99\textwidth]{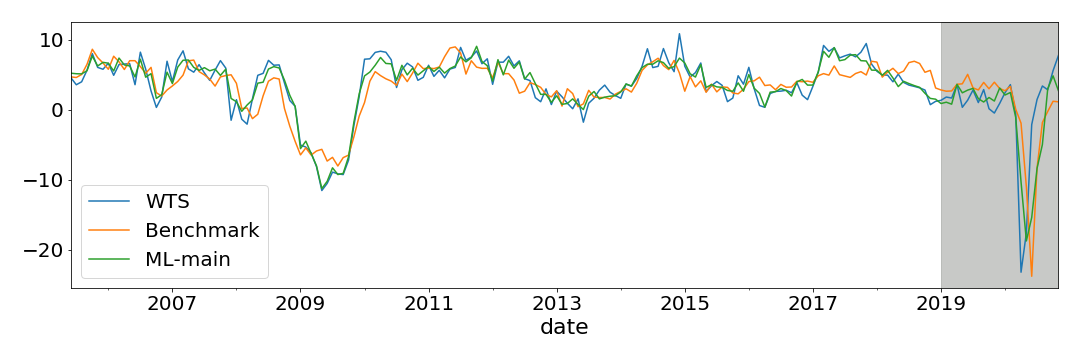}
 \caption{In-sample and out-of-sample prediction comparison for the ML main-case model (with lowest RMSE) and the benchmark model (OLS with base case) for time horizon $t+1$. The in-sample training period is Mar 2005 to Dec 2018, and the out-of-sample testing period is Jan 2019 to Dec 2020 (highlighted in gray).}
 \label{fig:testPredictions}
\end{figure}

Visual comparisons of the best performing ML model against the benchmark model for in-sample and out-of-sample (highlighted in gray) predictions are depicted in \autoref{fig:testPredictions}. Incorporating payments data in ML models provides downturn and recovery indications (during crisis periods) much better than the benchmark model in both in-sample and out-of-sample periods. 

\section{Nowcasting Performance for Normal and COVID-19 Periods}\label{apn:normalvscrisisre}

\begin{figure}[htbp]
 \centering
 \includegraphics[width=0.99\textwidth]{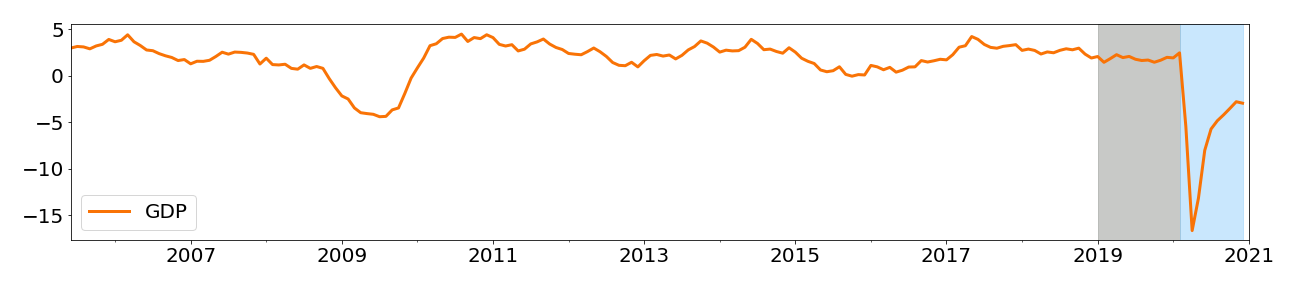}
  \caption{The test sample of GDP nowcasting exercises is divided into two sets: the pre-COVID-19 test set (highlighted in gray) and the COVID-19 test set (highlighted in blue).}
 \label{fig:gdp-norVscovid}
\end{figure}

In this section, we separately test our nowcasting model's out-of-sample performance during a normal time (Jan 19 to Feb 20) and the COVID-19 period (Mar 20 to Oct 20) of the test sample
highlighted in gray and blue, respectively, in~\autoref{fig:gdp-norVscovid}. To demonstrate, we use gradient boosting regression for these exercises. We observe a higher gain using payments data during the time of crisis (up to 35\% RMSE reduction) compared to the normal period of the test sample (15--25\% reduction in RMSE) using payments data (\autoref{tab:norvscovid}). These results demonstrate the usefulness of payments data during normal periods and crisis periods. 

\rowcolors{2}{gray!10}{white}
\begin{table}[htbp]
  \begin{center}
  \begin{threeparttable}
    \caption{Out-of-sample RMSE comparisons for seasonally adjusted YOY growth rates of GDP, RTS, and WTS at nowcasting horizon $t+1$ using the gradient boosting model\tnote{a}}
    \label{tab:norvscovid}
      \begin{tabular}{c c c c}
      \toprule
      \addlinespace
       \textbf{Targets} \ & \ \textbf{Pre-COVID-19 test set}\tnote{b} \ & \ \textbf{COVID-19 test set}\tnote{c} \\
      \addlinespace
      \bottomrule
         GDP  & 16 & 34 \tnote{} \\
         RTS  & 14 & 35 \tnote{} \\
         WTS  & 27 & 37 \tnote{} \\
      \bottomrule
      \end{tabular}
    \begin{tablenotes}\footnotesize
    \item [a] At time horizon $t+1$, we use current, i.e., $t$ month's payments data, to predict the same month's macro variables on the first day of the subsequent month.
    \item [b] For the pre-COVID-19 test set (or normal period): In-sample training period, Mar 2005 to Dec 2018, and out-of-sample testing period, Jan 2019 to Feb 2020.
    Those numbers show the percentage gain over benchmark cases for the same period. We use OLS with CPI, UNE, CFSI, CBCC, and the first available lagged target variable for the benchmark. 
    \item [c] For the COVID-19 test set (or crisis period): In-sample training period, Mar 2005 to Feb 2020, and out-of-sample testing period, Mar 2020 to Dec 2020. 
    These numbers show the percentage gain over benchmark cases for the same period.
    \end{tablenotes}
    \end{threeparttable}
  \end{center}
\end{table}

\section{Nowcasting Performance for First and Latest Vintages}\label{apn:realtimevslatest}

\begin{figure}[htbp]
 \centering
 \includegraphics[width=0.99\textwidth]{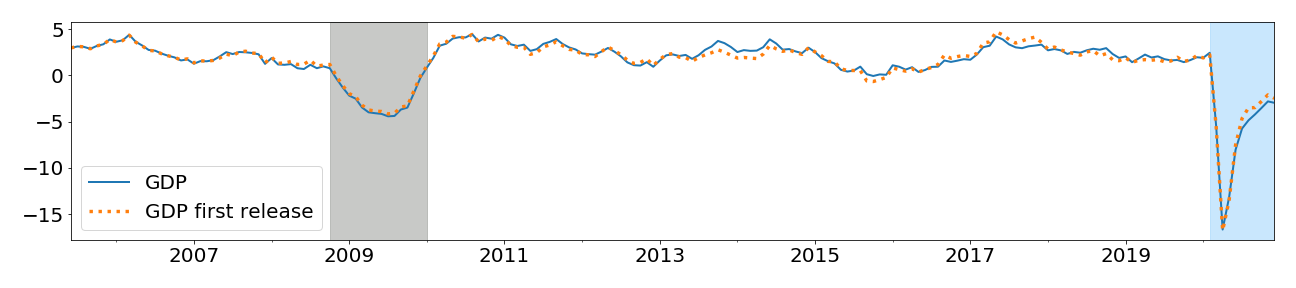}
  \caption{YOY seasonally adjusted GDP growth rates comparison of the first releases with latest releases. The GFC is highlighted in gray and the COVID-19 period is highlighted in blue.}
 \label{fig:gdp-firstVlatest}
\end{figure}

In this section, we compare the GDP nowcasting performance of our model with the real-time vintages (first releases) and the latest vintages (both shown in~\autoref{fig:gdp-firstVlatest}). Comparatively, the models using payments data perform better against the latest vintages (we get smaller RMSEs). However, the gains are small (\autoref{tab:realvslatest}). This makes sense given that the latest vintages are more accurate compared to the real-time vintages. Note: the performance gain is higher (about 10\%) at the nowcasting horizon $t+1$ compared to other time horizons. 

\rowcolors{2}{gray!10}{white}
\begin{table}[htbp]
  \begin{center}
  \begin{threeparttable}
    \caption{Out-of-sample RMSE comparisons for the seasonally adjusted YOY growth rate of GDP at nowcasting horizons $t$, $t+1$, and $t+2$ using the gradient boosting model\tnote{a}}
    \label{tab:realvslatest}
      \begin{tabular}{c c c c}
      \toprule
      \addlinespace
       \textbf{Nowcasting Horizon}\tnote{b} \ & \ \textbf{Latest Vintages}\tnote{c} \ & \ \textbf{Real-Time Vintages}\tnote{d} \\
      \addlinespace
      \bottomrule
         $t$   & 3.73 & 3.88\tnote{}  \\
         $t+1$ & 2.61 & 2.92\tnote{}  \\
         $t+2$ & 2.66 & 2.68\tnote{}  \\
      \bottomrule
      \end{tabular}
    \begin{tablenotes}\footnotesize
    \item [a] In-sample training period, Mar 2005 to Dec 2018, and out-of-sample testing period, Jan 2019 to Dec 2020. 
    \item [b] Nowcasting horizons: $t$ is on the first day of the month of interest (top panel), $t+1$ is on the first day after the month of interest (middle panel), and $t+2$ is on the first day, two months after the month of interest (bottom panel).
    \item [c] We use the latest available monthly levels of seasonally adjusted GDP from Statistics Canada Table 36-10-0434-01.
    \item [d] We use the historical real-time vintages (available as of Mar 2020) of adjusted monthly GDP from Statistics Canada Table 36-10-0491-01. 
    \end{tablenotes}
    \end{threeparttable}
  \end{center}
\end{table}

\end{document}